# Label-free mid-infrared photothermal microscopy revisits intracellular thermal dynamics: what do fluorescent nanothermometers measure?

Keiichiro Toda[1,2], Masaharu Takarada[3], Genki Ishigane[2], Hiroyuki Shimada[1], Venkata Ramaiah Badarla[1], Kohki Okabe[3] and Takuro Ideguchi[1,2,4,5,*]

[1] Institute for Photon Science and Technology, The University of Tokyo, Tokyo, Japan

[2] Department of Physics, The University of Tokyo, Tokyo, Japan

[3] Department of Pharmaceutical Science, The University of Tokyo, Tokyo, Japan

[4] Department of Advanced Materials Science, The University of Tokyo, Tokyo, Japan

[5] RIKEN Center for Advanced Photonics, RIKEN, Saitama, Japan

[*] Corresponding author: ideguchi@edu.k.u-tokyo.ac.jp

**Abstract**

Fluorescent nanothermometry has revealed pronounced intracellular temperature heterogeneity, establishing the field of single-cell thermal biology. However, these observations have sparked a controversy known as the "$10^5$ gap issue", because heat conduction calculations in aqueous environments predict that such large temperature distributions cannot be sustained within cells. Here, we address this issue using label-free mid-infrared photothermal microscopy. This technique quantifies heat-induced temperature changes under local thermal equilibrium (LTE), in accordance with the conventional thermodynamic and statistical-mechanical definition of temperature, by detecting refractive index variations. From transient thermal decay measurements, we determined that intracellular thermal diffusivity corresponds to 93-94% that of water. This result indicates that intracellular heat conduction is essentially water-like and rules out the hypothesis that anomalously slow intracellular heat conduction underlies the $10^5$ gap discrepancy. We then directly compared fluorescent nanothermometry with our label-free thermometry. Under seconds-long heating, the label-free method exhibited a rapid temperature response consistent with water-like heat conduction. In contrast, fluorescent nanothermometers showed not only a similarly fast response but also an additional slow variation that was absent in the label-free readout. This slow component cannot be explained by temperature changes defined under LTE, but instead likely reflects slower intracellular processes not governed by conductive heat transfer. These results suggest that the "$10^5$ gap issue" stems from comparing two fundamentally distinct physical quantities: the LTE-defined temperature and a slowly-varying, long-lived non-conductive signal detected by fluorescent nanothermometers. Our findings refine the description of intracellular temperature and open a new avenue for exploring previously unrecognized mechanisms of intracellular energy dynamics.

## Introduction

Over the past decade, various fluorescent nanothermometers have been developed to map intracellular temperature, and a central outcome of these efforts has been the observation of temperature heterogeneity of several kelvins within single cells[1,2,3,4,5]. Follow-up studies have further reported phenomena such as spatiotemporal thermoregulation in gene expression[6] and neural activity[7], as well as intracellular thermal signalling during neuronal differentiation[8]. These findings suggest that physiological processes may be thermally regulated at the cellular level and have given rise to the field of single-cell thermal biology[9,10]. Despite these advances, there remains a long-standing controversy known as the "$10^5$ gap issue"[11], which argues that such large temperature heterogeneity cannot exist within cells because water, the dominant cellular component (>70%), exhibits rapid heat conduction. Heat conduction calculations in aqueous environments predict that typical cellular heat production of 0.1-1 nW would generate temperature differences of only about $10^{-5}$ K, which is five orders of magnitude smaller than the experimentally reported values. Note that temperature in heat conduction calculation is defined under local thermal equilibrium (LTE) within the framework of thermodynamics and statistical mechanics (see "Photothermal mechamism underlying label-free thermometry" in Results for details).

There are two major scenarios to solve this issue. The first scenario is that intracellular heat conduction may be substantially slower than that in bulk water, leading to the large temperature heterogeneity. Specifically, a one- to two-order-of-magnitude reduction in thermal diffusivity, along with additional revisions to thermophysical parameters, has been proposed to account for the five-order discrepancy[12]. Several studies have attempted to quantify the intracellular thermal conduction properties; however, reported values vary substantially (Table 1). For example, studies using fluorescent nanothermometry reported that intracellular heat conduction is nearly an order of magnitude slower than that of water (18-19% of water)[13,14], whereas a label-free coherent Raman scattering approach reported water-like heat conduction (89-93% of water)[15]. Although the Raman-based method mitigates systematic uncertainties associated with fluorescent probes (as discussed later in this work), its limited temporal resolution prevents direct observation of transient heat diffusion on the microsecond timescale. Instead, thermal conductivity was inferred from steady-state temperature rises under continuous heating. Measurements performed at such long timescales may be influenced by biological responses, including heat-induced changes in specific heat capacity and density, which can compromise the accuracy of the inferred conductivity (see Discussion for details). Recently, transient thermal decay was examined using a semi-label-free approach based on dark-field detection of the photothermal effect from gold nanoparticles (GNPs) that served as localized heat sources[16]. In that study, decay dynamics were comparably evaluated in water and intracellular environments, showing slower decay in cells (51-53±15-20% of water). However, the decay time reported for a 100-nm heat source in water was unrealistically long, exceeding 100 μs, suggesting the presence of systematic errors, likely arising from a lack of low-spatial-frequency information inherent to dark-field imaging (see Discussion for details). These substantial discrepancies across methodologies underscore the need for carefully designed, label-free measurements that directly quantify microsecond-scale heat diffusion dynamics, enabling accurate determination of intracellular thermal conduction properties on firm quantitative grounds.

The second scenario questions whether fluorescent nanothermometers measure temperature defined under LTE or instead respond to other intracellular parameters. In this context, extensive control experiments have excluded major confounding factors, including probe sensitivity to pH or viscosity[1,2,3,4]. In addition, similar temperature distributions have been observed with probes based on different sensing mechanisms, suggesting that these sensors detect a physical quantity closely associated with temperature. Nevertheless, definitive experimental evidence demonstrating that these readouts correspond to LTE-defined temperature is still lacking, as no study has directly compared fluorescent nanothermometry with an independent reference method that quantifies temperature defined under LTE. Such validation requires a reference thermometer that operates under identical conditions within the same microscopy platform. More importantly, the reference method must be grounded in a clearly defined physical principle that directly links the measured observable to the LTE-defined temperature.

Taken together, the $10^5$-gap controversy can be reduced to two fundamental questions: whether intracellular heat conduction is substantially slower than in water, and whether fluorescent nanothermometer readouts faithfully report temperature defined under LTE. In this study, we address both questions using mid-infrared photothermal optical diffraction tomography (MIP-ODT)[17], a label-free thermometry technique. MIP-ODT detects refractive index (RI) changes induced by local heating of water molecules via mid-infrared vibrational absorption. Since the RI variation primarily arises from density changes associated with thermal expansion, the measured signal directly reflects temperature changes defined under LTE within the diffraction-limited observation volume. This LTE-defined temperature readout enables quantitative determination of intracellular thermal conduction properties and provides an independent reference for evaluating fluorescent nanothermometer readouts. MIP-ODT offers two features that are essential for accurate quantification of intracellular heat conduction. First, by operating in a pump-probe imaging mode with nanosecond impulsive heating and time-resolved detection, we can directly measure transient thermal decay on the microsecond timescale, which minimizes the influence of slower intracellular processes. Second, unlike dark-field photothermal measurements[16], ODT preserves low-spatial-frequency information required for quantitative reconstruction of the heat distribution. These capabilities enable direct and quantitative measurement of intracellular thermal dynamics under well-defined LTE conditions. Using this approach, we determine intracellular thermal diffusivity in living cells and compare fluorescent nanothermometer readouts against an LTE-defined temperature reference, thereby clarifying the origin of the $10^5$ discrepancy.

## Results

### Photothermal mechanism underlying label-free thermometry.

To measure temperature change defined under LTE, we exploit the photothermal response induced by infrared (IR) absorption. Figure 1a illustrates the spatiotemporal sequence of photothermal processes triggered by IR absorption. Upon IR absorption, resonant vibrational modes in specific molecules are selectively excited. The excited vibrations decay within ~10 ps under cellular conditions and their energy is rapidly redistributed into surrounding molecular degrees of freedom, including vibrational, rotational, and translational motions. This ultrafast energy redistribution rapidly equilibrates the system and establishes LTE within a local region. The ensemble of rapidly equilibrating molecular degrees of freedom follows a Boltzmann distribution characterized by a single temperature. LTE is generally expected to be well defined on length scales of ~10 nm or larger[18], with corresponding equilibrium times on the order of nanoseconds. Subsequently, at single-cell length scales, the LTE-defined temperature evolves through heat conduction on the timescale of microseconds to milliseconds. The resulting temperature change can be detected as RI variations arising from density changes associated with thermal expansion, which also occur on the nanosecond timescale.

It is important to note that these RI variations reflect the averaged thermodynamic response of the molecular ensemble within the diffraction-limited observation volume and therefore do not selectively probe individual molecular excitations or specific energy modes. This ensemble-averaged nature enables robust quantification of LTE-defined temperature, even in potential cases where certain internal molecular degrees of freedom do not equilibrate within the measurement window. For example, slowly relaxing modes, such as biomolecular conformational transitions, may remain out of equilibrium. However, the temporarily stored energy in such modes does not contribute to the LTE-defined temperature and therefore does not drive the thermal-expansion-mediated RI change. Consequently, our label-free photothermal approach provides a more robust readout of LTE-defined temperature than fluorescent nanothermometry, which relies on nanometer-localized probes whose fluorescence can be influenced by the surrounding nanoscale molecular environment, including local non-equilibrium energy states.

**MIP-ODT system.**

We implement label-free thermometry using mid-infrared photothermal optical diffraction tomography (MIP-ODT)[17], a recently developed label-free chemical imaging technique. A conceptual schematic is shown in Fig. 1b, and system details are provided in "MIP-ODT system" in Methods. In this method, a focused IR laser beam tuned to vibrational absorption band of water generates a micrometer-scale artificial heat source inside the cell. Because water is distributed nearly uniformly throughout the cell, this approach produces a heat source with a well-determined spatial extent. Precise control of the heat-spot size is essential for quantitative thermal analysis, as the decay timescale of heat diffusion strongly depends on the dimensions of the heat source. To confine heating intracellularly, we carefully select the IR wavelength (wavenumber) such that the IR light is dominantly absorbed within a cell (77-99% depending on the cellular thickness), thereby ensuring localized intracellular heating (see "Selection of IR wavenumber" in Methods). The resulting photothermal response is detected as heat-induced RI changes using ODT implemented with off-axis digital holography (DH)[19] and azimuthally scanned illumination. This configuration reconstructs three-dimensional (3D) RI volumetric images with lateral and axial resolutions of 250 nm and 5 µm, respectively.

Subtracting RI images measured with and without IR illumination yields the heat-induced RI change, $\Delta n$ (Fig. 1c). In addition to the rapid RI change associated with thermal expansion, $\Delta n$ can include a slower component that reflects heat-driven redistribution of macrobiomolecules, primarily governed by thermophoresis and subsequent molecular diffusion[20]. Under prolonged heating, this redistribution develops over milliseconds to seconds[20-22], whereas the thermal-expansion response evolves on microsecond timescales through heat conduction. To isolate the thermal-expansion component only, we perform pump-probe imaging on the timescale of microseconds with nanosecond impulsive heating (Fig. 1d). We define the RI change induced by impulsive heating as $\delta n_\rho$, which is expressed as

$$\delta n_\rho = \left(\frac{dn}{dT}\right)_{T=T_0} \delta T(t), \quad \text{(Eq. 1)}$$

where $\frac{dn}{dT}$ is the thermo-optic coefficient at the initial temperature $T_0$, and $\delta T(t)$ is the transient temperature change induced by impulsive heating. Although $\frac{dn}{dT}$ is intrinsically temperature dependent, it can be treated as constant in the pump-probe measurement because the temperature relaxes back to $T_0$ due to heat diffusion before the arrival of the subsequent pump pulse. Consequently, time-resolved measurements of $\delta n_\rho$ directly report the temporal evolution of $\delta T$ governed by heat conduction.

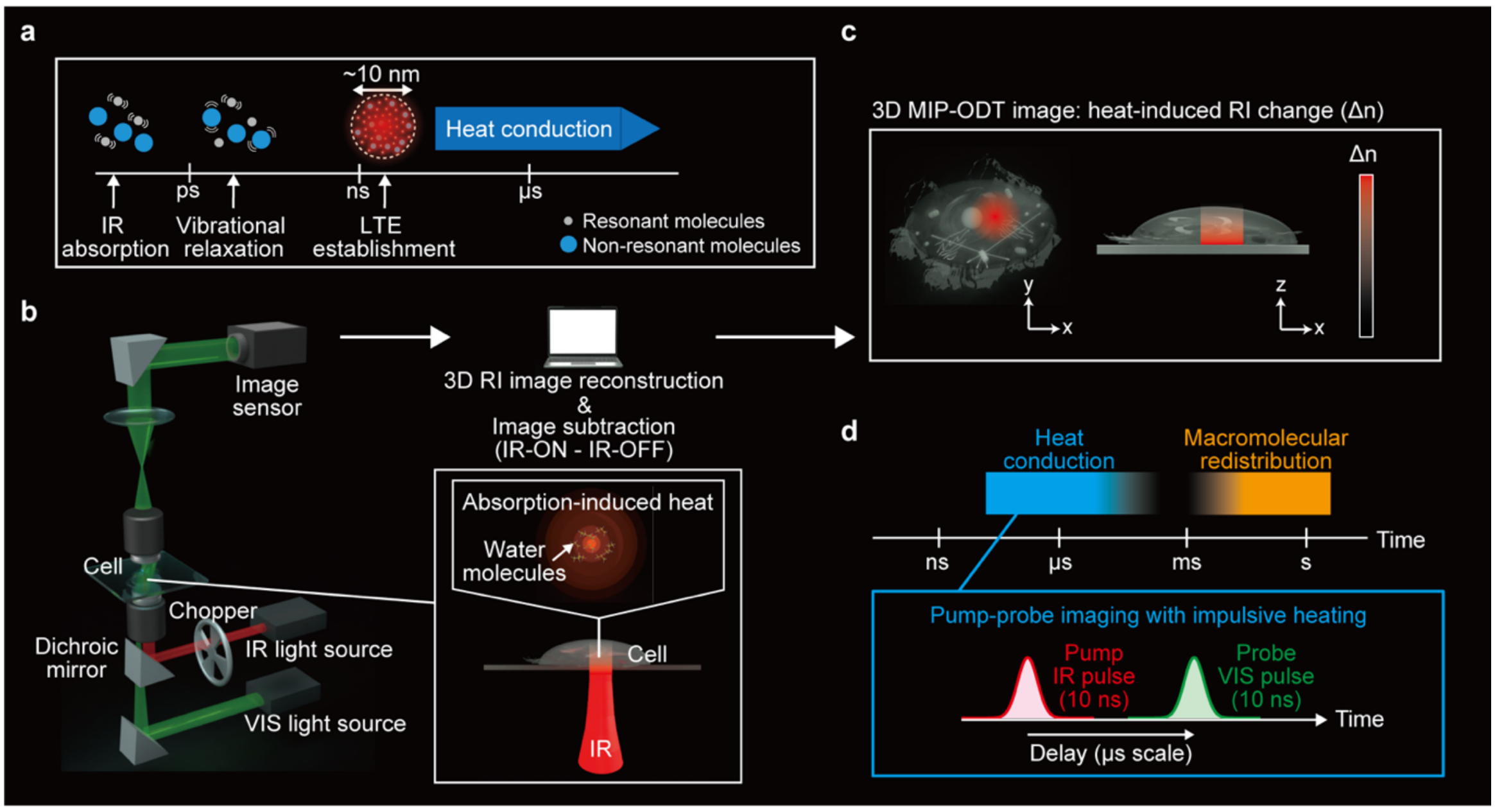

**Fig. 1 Principle and schematic of MIP-ODT for label-free thermometry. a** Spatiotemporal evolution of energy relaxation and equilibration following IR excitation. **b** Schematic of the MIP-ODT system. **c** Illustration of a 3D MIP-ODT image. **d** Timescales of heat conduction and marcomolecular redistribution. Heat conduction can be selectively measured by pump-probe imaging with nanosecond pulses.

**Intracellular thermal diffusivity measurement.**

To evaluate intracellular thermal diffusivity, we performed transient heat diffusion measurements by pump-probe imaging with 10-ns impulsive IR heating and visible detection pulses. The delay between the IR and the visible pulses is scanned at 1 μs step so that the thermal decay can be well visualized (see "Thermal diffusivity measurement" in Methods for details). For system evaluation, we first measured water and glycerol as reference samples. The wavenumber of the IR light was set to vibrational resonances of water (3,150 $cm^{-1}$) and glycerol (2,925 $cm^{-1}$), and a ~5-nJ pulse was illuminated onto the sample with a full-width at half maximum (FWHM) spot size of ~3.5 μm. Figure 2a shows $\delta T$ images of water captured at delays of 3, 7, and 13 μs after the impulsive heating, which are converted from raw images of $\delta n_\rho$ by an experimentally derived nearly linear calibration curve (see "Calibration curve of $\delta n_\rho$ on $\delta T$" in Methods for details). From the transient evolution of the $\delta T$ map, we plotted its decay (Fig. 2b) by taking the average values of an integration volume at the center part (1.8×1.8 µm) of the $\delta T$ profile in the x-y plane, indicated by a white dotted area in Fig. 2a.

To determine the thermal diffusivity from the measured data, one must carefully consider the spatial profile of the temperature map because heat conduction strongly depends on the initial spatial profile of the heat source and cannot be modeled by a single exponential function (see Supplementary Note 7 for details). Therefore, we performed heat conduction calculations to find a thermal diffusivity that replicates the experimental data (See "Heat conduction calculation" in Methods for details). The thermal diffusivity was determined by minimizing the sum of the squares of the differences between the experimental and simulation data points with a resolution of $0.001\times10^{-6}$ $m^2s^{-1}$.

The thermal diffusivities of water and glycerol obtained by the aforementioned analysis are $(0.143\pm0.004)\times10^{-6}$ $m^2s^{-1}$ and $(0.089\pm0.002)\times10^{-6}$ $m^2s^{-1}$ for eight samples, respectively (see Fig. 2d), and are in good agreement with literature values, $0.144\times10^{-6}$ $m^2s^{-1}$ and $0.092\times10^{-6}$ $m^2s^{-1}$ within 2% deviation[23]. Furthermore, the ratio of the standard deviation (STD) to the mean values is 2.7% for the water sample, which is lower than those reported in thermal conductivity measurements of previous studies (7.1%[16] (GNP) and ~40%[13] (NVC)). This demonstration verifies our system's ability to measure heat conduction parameters with higher precision than previous works.

Then, we measured the thermal diffusivity of the cytoplasm and nucleus in living COS7 cells by selectively heating intracellular regions, as illustrated in Fig. 2c (see "Sample preparations" in Methods for sample preparation procedure). Note that we have confirmed the slight IR absorption by surrounding water along the depth axis has only a minor impact on the measured intracellular thermal diffusivity (see "Thermal diffusivity measurement" in Methods). The measured thermal diffusivities of the cytoplasm and nucleus are $(0.133\pm0.002)\times10^{-6}$ $m^2s^{-1}$ and $(0.134\pm0.008)\times10^{-6}$ $m^2s^{-1}$ for 8 and 9 samples, respectively (~93-94 % against that of water). The corresponding thermal conductivity, $\kappa = \alpha/c\rho$, is calculated as 0.56 $Wm^{-1}K^{-1}$ for both the nucleus and cytoplasm by using literature values of a cell density $\rho$ of 1.08 kg $m^{-3}$ [24] and a specific heat capacity $c$ of 3,900 J $kg^{-1}K^{-1}$ [25]. It is consistent with theoretically estimated values (~0.54 $Wm^{-1}K^{-1}$) by Levy's model and the effective medium theory model by assuming the proportion of intracellular constituent molecules as water (80%) and protein (20%)[26], where the protein's thermal conductivity is referenced by a reported value (0.27 $Wm^{-1}K^{-1}$)[27]. Consequently, our measurements suggest that intracellular heat conduction is predominantly determined by the constituent molecular species rather than the intracellular structures, and it is likely that water, the most abundant intracellular molecular species, dominates micrometer-scale intracellular thermal conduction.

Table 1 presents a comparison of our work with previous investigations. Our results significantly differ from thermal diffusivity obtained with a GFP fluorescent thermometer ($0.027\times10^{-6}$ $m^2s^{-1}$)[14] and thermal conductivities obtained with NVCs (0.11 $Wm^{-1}K^{-1}$)[13], and GNP plasmonic imaging (0.31-0.32 $Wm^{-1}K^{-1}$)[16], while similar to those with coherent Raman scattering (CRS) microscopy (0.563 $Wm^{-1}K^{-1}$, 0.587 $Wm^{-1}K^{-1}$, fixed cell)[15] and heat-flow measurements operated outside a cell (0.54-0.57 $Wm^{-1}K^{-1}$)[26,28,29]. We will discuss the reasons for the discrepancy in the discussion section. Our result of water-like thermal diffusivity in living cells contradicts the hypothesis that intracellular heat conduction is considerably slower than in water.

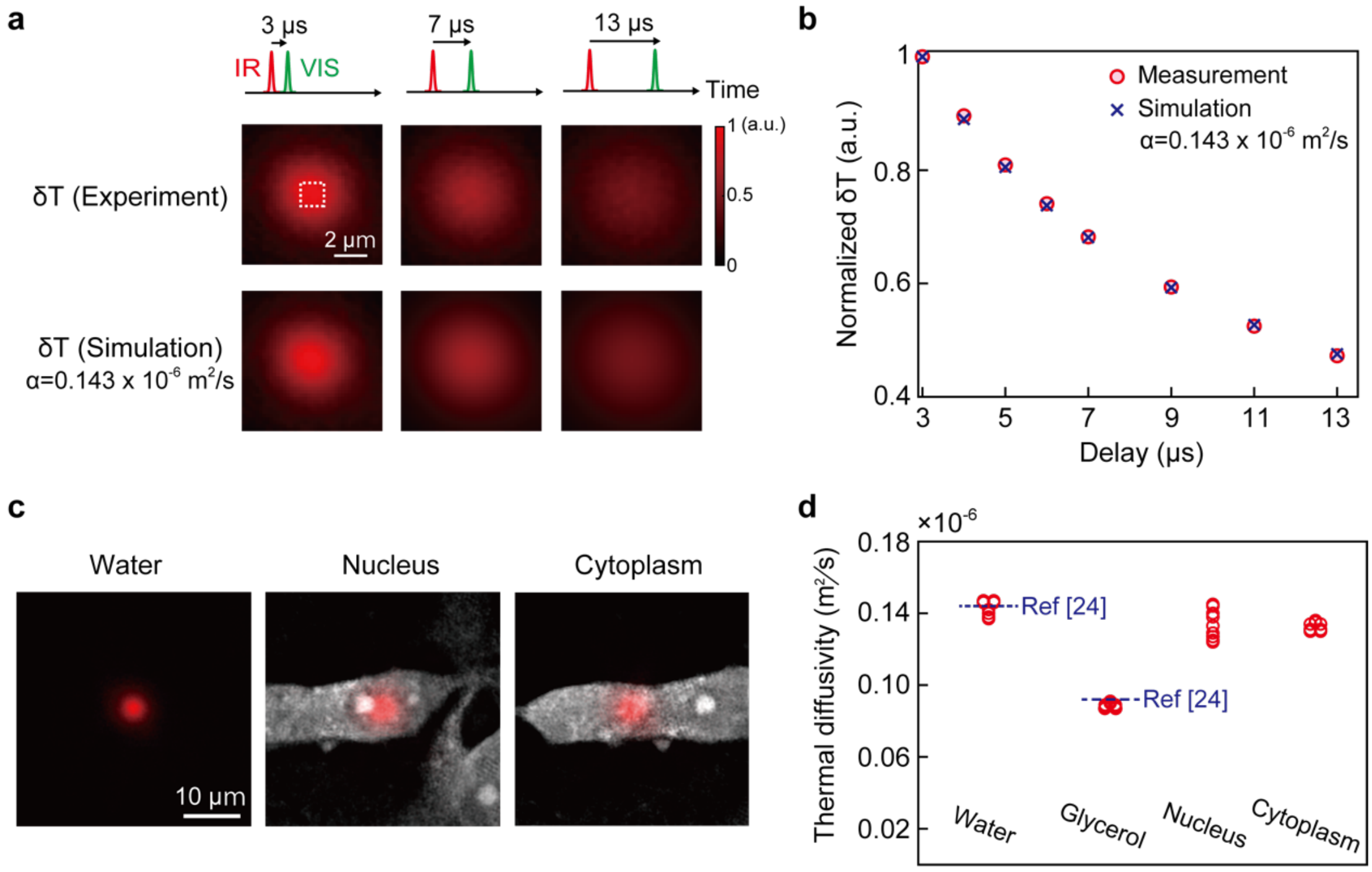

**Fig. 2 Intracellular thermal diffusivity measurement by pump-probe imaging with 10-ns impulsive heating. a** Experimentally measured temperature change ($\delta T$) maps induced by a nanosecond IR heating (top) and simulated $\delta T$ maps with a thermal diffusivity of $0.143\times10^{-6}$ $m^2s^{-1}$ (bottom) at delays of 3, 7, 13 μs. **b** Normalized $\delta T$ as a function of delay. The plotted data are averages within the center part of the $\delta T$ map shown as the white dotted area in **a**. **c** Images of the IR heating spot (shown in red) in water, nucleus, and cytoplasm of a COS7 cell. **d** Measured thermal diffusivities for water, glycerol, nucleus, and cytoplasm. The dotted lines represent the literature values of thermal diffusivities for water and glycerol[24].

**Table 1 Comparison of thermal conduction parameters measured in this work and previous works.**

| **Thermometer** | **Intracellular thermal diffusivity ($10^{-6}$ $m^2s^{-1}$)** (/water (%)) | **Intracellular thermal conductivity ($Wm^{-1}K^{-1}$)** (/water (%)) | **STD of thermal conductivity for water** | **Cell type Temperature** |
|---|---|---|---|---|
| **GFP**[15] Fluorescence intensity | 0.027 (19%) | N/A | N/A | HeLa 37 °C |
| **NVCs**[14] Fluorescence intensity | N/A | 0.11 (18 ± ~6.6%) | ±~40%** | HeLa 27 °C |
| **GNP plasmonic**[18] RI change | N/A | 0.31-0.32 (51-53 ± 15-20%**) | ±7.1% | HeLa, MCF-7, etc. 25 °C |
| **CRS**[16] Spectral shape of OH bonds | N/A | 0.563 (nucleus) (89%) 0.587 (cytoplasm) (93%) | N/A | A549 (fixed) 37 °C |
| **MIP-ODT (this work)** RI change | 0.134 (94 ± 1.7%) 0.133 (93 ± 5.6%) | 0.56* (nucleus) (92%) 0.56* (cytoplasm) (92%) | ±2.7% | COS7 24 °C |

* These values represent the thermal conductivities estimated from measured thermal diffusivities with the literature values of heat capacity[25] and density[24].

** These are estimated values from the graphs shown in the papers.

**Comparison of MIP-ODT thermometry with fluorescent nanothermomatry.**

We compared our MIP-ODT thermometry and fluorescent nanothermometry under the identical experimental conditions within the same microscopy setup. To enable this comparison, we integrated a fluorescence imaging branch into the MIP-ODT system (See "Fluorescence imaging" in Methods for details). Pump-probe imaging on the microsecond timescale using impulsive nanosecond heating is not suitable for this comparison because fluorescent nanothermometers typically respond on the millisecond timescale[30]. We therefore applied seconds-long continuous heating and monitored the resulting temperature changes. Continuous heating was achieved by focusing a 1456-nm CW-IR laser, resonant with a water overtone band, onto the sample with a 5 μm FWHM spot size. Images were acquired at a temporal resolution of 10 ms, determined by the frame rate of the image sensor (See "Heat-induced intracellular temperature change measurement" in Methods for details).

Under second-long continuous heating, however, $\Delta n$ measured by our MIP-ODT contains not only the rapid thermal-expansion response but also a slow RI change induced by molecular redistribution described in "MIP-ODT system" section of Results. This additional contribution prevents direct isolation of the temperature component (See Supplementary Note 11 for details). To track the temporal evolution of the temperature change induced by continuous heating, $\Delta T(t)$, we developed a dual-heating method in which nanosecond impulsive heating is superimposed on continuous heating (See "Dual-heating method" in Methods for details). A timing diagram of this measurement is shown in Fig. 3a. In this approach, the RI change induced solely by the impulsive heating ($\delta n_\rho$) is obtained by subtracting images acquired with and without impulsive heating. Importantly, the impulsive heating is modulated ON and OFF at 50 Hz, which is much faster than the characteristic evolution time of molecular redistribution. As a result, the change in the slow component is negligible between the two measurements and is canceled by subtraction. The extracted signal is given by

$$\delta n_\rho = \left(\frac{dn}{dT}\right)_{T=T_0+\Delta T(t)} \delta T, \qquad \text{(Eq. 2)}$$

where $T_0$ is the initial temperature and $\delta T$ is the temperature change induced by impulsive heating. In this measurement, $\delta T$ is constant because the pump-probe delay is fixed. Since the thermo-optic coefficient $\frac{dn}{dT}$ depends on temperature, the continuous-heating-induced temperature change $\Delta T(t)$ can be extracted from the measured $\delta n_\rho$. We validated the temperature measurement capability of this approach by tracking the dynamic temperature response in a water sample (see Supplementary Note 10 for details).

Figure 3b shows the temporal evolution of intracellular $\Delta n$ with 2.4-s continuous heating. Brown crosses and black squares represent $\Delta n$ measured in frames with impulsive heating ON and OFF, respectively. The red circles at the bottom of the figure present their difference ($\delta n_\rho$). Using the temperature dependence of the thermo-optic coefficient, $\delta n_\rho$ was converted into $\Delta T$ (See "Calibration curve of $\delta n_\rho$ on $\Delta T$" in Methods for details). The extracted $\Delta T$ increased within the first frame interval of 20 ms and then remained constant at ~5.0 K. This behavior is consistent with heat conduction calculations based on water-like intracellular thermal diffusivity, which predict a rise time of ~100 µs. Similar $\Delta T(t)$ profiles were observed in other cells, even when $\Delta n$ exhibited substantial variability due to redistribution-related RI contributions with distinct temporal dynamics. This consistency demonstrates robust cancellation of the redistribution-related component by the dual-heating subtraction scheme (see Supplementary Note 17 for details).

We next measured $\Delta T$ using fluorescent polymeric thermometer (FPT) and directly compared the results with those obtained by MIP-ODT thermometry. Because continuous heating can induce thermophoresis, which alters probe concentration, we employed ratiometric FPT thermometry. In this approach, temperature is determined from the ratio of fluorescence intensities at two wavelengths, thereby canceling signal variations arising from probe redistribution[31].

The top panel of Fig. 3c shows $\Delta T$ measured by FPT and reveals two distinct features. These features were consistently observed aross different heating locations and samples. The first feature is a rapid rise and fall at the onset and termination of the heating, with a response time shorter than the 40 ms frame interval. A comparable fast response is observed in the MIP-ODT measurement shown in the bottom panel of Fig. 3c. The temperature changes measured by the two methods are similar, ~5 K for MIP-ODT and 4 K for FPT thermometry. This agreement indicate that, on short timescales, FPT reports temperature changes consistent with heat conduction. The second feature is a gradual increase in $\Delta T$ during continuous heating, followed by slow relaxation after heating ceases. This behavior occurs on a timescale of seconds and is not observed in the MIP-ODT measurement. Its absence in MIP-ODT readout incdicates that the slowly varying signal does not correspond to temperature defined under LTE.

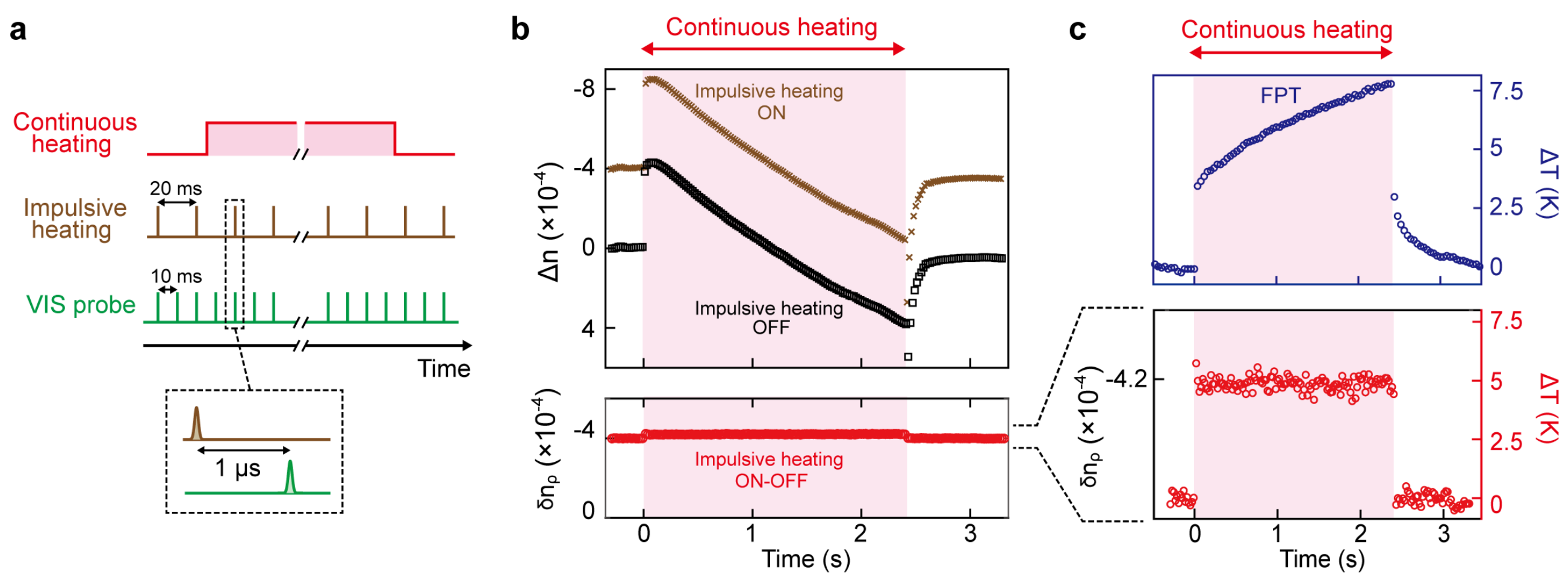

**Fig. 3 Heat-induced intracellular temperature change measurements by MIP-ODT thermometry and fluorescent nanothermometry. a**. A timing chart of the continuous heating, impulsive heating, and visible probe for temperature change measurement. **b** A temporal evolution of $\Delta n$ under continuous heating with (brown crosses) and without (black squares) the impulsive heating, and their difference, $\delta n_\rho$ (red circles). **c** Time variation of $\Delta T$

induced by continuous heating measured by fluorescent nanothermometer (FPT) (top) and label-free MIP-ODT thermometer (bottom).

We obtained multiple lines of evidence indicating that the slowly varying component is unlikely to arise from methodological artifacts of fluorescent nanothermometry but instead reflects an intracellular phenomenon. First, we reproducibly observed a comparable slow component in cells using a different fluorescent probe, Rhodamine B, which relies on a different temperature-sensing mechanism (Fig. 4a; see Methods, "Investigation of potential artifacts in fluorescent nanothermometry," for experimental details). The persistence of this signal across probes operating on different physical principles argues against probe-specific artifacts. Second, we found that the slow signal was absent in liposome-encapsulated water (Fig. 4b). This non-cellular control measurement excludes experimental artifacts, including direct IR excitation of the probe, as the origin of the signal, and indicates that the signal depends on the intracellular environment. These cross-probe comparisons and control experiments support the interpretation that the slowly varying component arises from an intracellular phenomenon.

Additional support for this interpretation is provided by an independent study conducted by our co-authors, currently under peer review[30]. That study systematically excluded several potential artifacts, including direct IR excitation of intracellular biomolecules, delayed probe response kinetics within cells, and diffusion-driven probe redistribution, as plausible origins of the slow signal (see Supplementary Note 18). Importantly, whereas their study narrowed the possible origins of this signal by systematically excluding probe-intrinsic artifacts, our study evaluates the fluorescence readout against a LTE-defined temperature reference measured by MIP-ODT. These two studies thus address the slow component using distinct experimental strategies.

We next investigated how this slow component is sensed by the fluorescent probe. Specifically, we tested whether the slow component is detected through the same thermoresponsive unit that senses the LTE-defined temperature. To address this question, we developed a control copolymer (CP) that retains the chemical backbone of FPT but incorporates a slightly modified thermoresponsive unit[32]. In this design, the amide side chain of the thermoresponsive unit was modified from an ethyl group to an isopropyl group, which is expected to significantly reduce the temperature-dependent fluorescence change between 20 °C and 40 °C. When examined under experimental conditions identical to those used in Fig. 4a, the CP exhibited both a reduced LTE-defined temperature signal and a markedly attenuated slow component (Fig. 4c). The attenuation of the slow component upon modification of the thermoresponsive unit indicates that the slow signal is sensed through this unit, as is the LTE-defined temperature response. This finding suggests that the slow component reflects an intracellular process that is sensed through the molecular transition responsible for LTE-defined temperature detection.

Finally, we sought to further constrain the intracellular processes underlying the slow component. To test whether the slow component is associated with major metabolic pathways, we pharmacologically inhibited glycolysis and oxidative phosphorylation. These interventions did not abolish the slow signal (Supplementary Note 19), indicating that the slow component is unlikely to originate from processes directly coupled to these major metabolic pathways.

In contrast, ref. 30 reports that the slow component is absent in giant plasma membrane vesicles, which retain some cytoplasmic constituents while lacking intact intracellular organization and metabolic activity. These observations suggest that the slow component may reflect intracellular processes that require intact cellular organization and biomolecular assemblies. Notably, this slow component emerges only under seconds-scale heating and reaches a steady state within approximately 5 to 10 s[30]. A detailed discussion of its physical origin and implications is provided in the Discussion section.

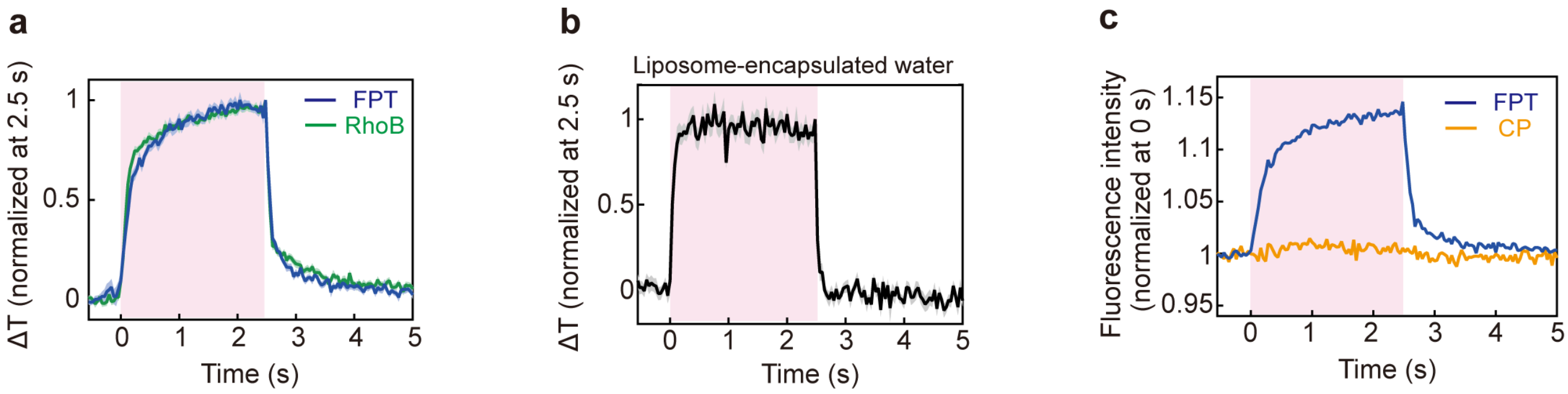

**Fig. 4 Heat-induced intracellular temperature responses measured by fluorescent nanothermometry. a** Comparison of normalized temperature readouts ($\Delta T$) under seconds-long continuous heating measured with two fluorescent nanothermometers: FPT (blue) and Rhodamine B (RhoB, green). **b** Normalized $\Delta T$ response measured with RhoB in liposome-encapsulated water under the same continuous-heating condition. For **a** and **b**, traces are normalized to the value at $t = 2.5$ s; shaded regions indicate s.e.m. (n = 30 independent measurements). **c** Normalized fluorescence-intensity response to continuous heating measured by FPT (blue) and a control polymer (CP) (orange) that retains the same chemical backbone as FPT but incorporates a slightly modified thermoresponsive unit designed to have substantially reduced temperature sensitivity.

## Discussion

We first discuss the initial part of this study, the measurement of intracellular thermal diffusivity. To our knowledge, this study provides the first label-free determination of thermal diffusivity in living cells. We therefore compare the thermal conductivity evaluated in this work with the previous label-free studies[15,16], instead. There is a large discrepancy between the values determined in this work (0.56 $Wm^{-1}K^{-1}$) and the previous dark-field photothermal study (0.31-0.32 $Wm^{-1}K^{-1}$)[16], although both approaches detect the photothermal signals. A plausible explanation for this difference lies in the intrinsic spatial high-pass filtering effect of dark-field imaging, which could induce a systematic error. In that study, a decay time over 100 μs was reported for a 100-nm heat source in water, whereas theoretical estimates predict sub-microsecond decay. We suspect that the insufficient temporal resolution of 5 μs led to measurement of the decay tail after substantial heat spreading had already occurred. Although a small spot was observed at the beginning of the diffusion in their images, this feature may be due to the intrinsic spatial filtering nature of dark-field imaging, which causes a lack of low-spatial-frequency information and can artificially visualize the vastly spreading heat distribution as a small spot. In contrast, MIP-ODT does not suffer from the spatial filtering effect, and it was experimentally verified by comparing measured and calculated results in RI signal decay in water and glycerol.

Our result (0.56 $Wm^{-1}K^{-1}$ in nucleus and cytoplasm) closely matches values reported using CRS microscopy (0.563 and 0.587 $Wm^{-1}K^{-1}$ in nucleus and cytoplasm, respectively), obtained from steady-state temperature distributions in fixed cells[15]. However, the methodology used in that work may be susceptible to inaccuracies arising from slowly varying parameters. Their analysis relied on the steady-state heat conduction equation, $\kappa\nabla^2 T = -Q$, where $\kappa$, $T$, and $Q$ represent thermal conductivity, temperature, and the amount of heat supplied per unit time, respectively. Since $\kappa$ can be expressed as $\kappa = c\rho\alpha$, where $c$, $\rho$, $\alpha$ represent specific heat capacity, density, and thermal diffusivity, respectively, slow heat-induced variations in the parameters ($c$ or $\rho$), as well as other factors not explicitly accounted for in the equation, could affect the inferred conductivity. Indeed, a study has shown that the specific heat capacity $c$ of living cells can change by tens of percent during seconds-long heating[33]. A likely reason why the steady-state measurement nevertheless yielded accurate intracellular thermal conductivity without suffering from slowly varying effects is that the chemical fixation prevented these processes. In contrast, our approach determines thermal diffusivity directly from transient decay dynamics on the microsecond timescale, described by $\partial T/\partial t = \alpha\nabla^2 T$ (where $t$ denotes time). This transient formulation renders the measurement insensitive to slower variations in other intracellular parameters, even in living cells. It is also important to note that the sample and temperature (fixed A549 cells under 37°C) are different from ours (living COS7 cells under 24°C), implying that thermal conductivity does not strongly depend on cell types and temperature.

We next discuss the latter part of this study, the comparison between fluorescent nanothermometry and MIP-ODT thermometry, and investigate the origin of the slowly varying component. Our measurements show that this component does not correspond to temperature defined under LTE and therefore does not follow the temporal profile predicted by conductive heat transfer. Instead, it evolves on timescales of seconds. When considered together with the control experiments, these findings suggest that the slowly varying component is associated with intracellular processes that relax more slowly than conductive heat dissipation. The control experiments further indicate that the slow component is detected through molecular transition involved in LTE-defined temperature sensing by fluorescent nanothermometers and likely requires intact cellular organization and biomolecular assemblies. These characteristics are consistent with slow-relaxing intracellular processes involving conformational or structural rearrangements. Possible examples include structural transitions of RNA[34] and mechanically mediated energy redistribution along the cytoskeleton network[35], both of which represent nanoscale non-equilibrium intracellular processes occurring on timescales of seconds. This interpretation is also consistent with the different measurement principles of fluorescent nanothermometry and MIP-ODT. MIP-ODT reports the ensemble-averaged LTE-defined temperature, whereas fluorescent nanothermometers can also be influenced by energy states in the surrounding nanoscale environment associated with such intracellular processes.

Based on this interpretation of the slow component, the coexistence of a rapidly equilibrating LTE-defined temperature response and a slower non-conductive intracellular component motivates a description within a non-equilibrium thermodynamic framework. In this framework, the local thermodynamic state is not fully characterized by a single temperature $T$, but instead requires an additional state variable $\xi$, an internal variable that could be associated with slow conformational or structural modes. In a minimal formulation, the internal energy density is expressed as

$$u(T, \xi) = u_F(T) + u_S(\xi), \quad \text{(Eq. 3)}$$

where $u_F(T)$ represents the energy stored in rapidly equilibrating degrees of freedom that establish LTE and govern conductive heat transfer. By contrast, $u_S(\xi)$ represents energy stored in slower internal modes whose relaxation dynamics are decoupled from heat conduction and may not reach equilibrium within our experimental observation window (i.e., 2.5 s). Although optical heating initially perturbs only the rapidly equilibrating degrees of freedom that define $T$, the observed slowly varying signal suggests that energy is subsequently transferred to slower internal modes, thereby driving the evolution of the internal variable $\xi$. This energy exchange may underlie the slowly varying signal detected by fluorescent nanothermometers. If the slow modes reach partial internal equilibration, their state may be parameterized by an effective temperature $T_{\text{eff}}$ ($\neq T$). Importantly, this effective temperature is conceptually distinct from the LTE-defined temperature that determines conductive heat transfer.

Finally, we address and provide comments on the controversial issues in the field of cellular thermal biology. First, the existence of the non-conductive component suggests that measurements of thermal conductivity or diffusivity using fluorescent nanothermometry may be strongly influenced by non-conductive intracellular responses, rather than reflecting conductive heat transfer alone. Therefore, fluorescence readouts cannot be regarded as intrinsic measures of conductive thermal dynamics. Second, in the context of the $10^5$ gap issue, the large discrepancy may arise from a comparison between two fundamentally distinct physical quantities: LTE-defined temperature, which evolves through conductive heat transfer, and a long-lived non-conductive intracellular signal detected by fluorescent nanothermometers. Because these quantities follow different relaxation dynamics, direct comparison can lead to a substantial discrepancy. In our measurements, the non-conductive signal decays on a timescale of ~1 s, whereas conductive heat dissipation occurrs within ~10 $\mu$s. This five-order-of-magnitude difference in relaxation times implies that the slowly relaxing component persists and accumulates over time. As a result, its integrated response can appear approximately five orders of magnitude larger, potentially accounting for a $10^5$ enhancement observed in fluorescence-based readouts.

## Conclusion

In this study, we addressed the $10^5$ gap issue using label-free MIP-ODT, which quantifies heat-induced temperature changes under LTE by measuring RI variations. We first determined intracellular thermal diffusivity from transient thermal-decay measurements. The method was validated using water and glycerol as reference materials, yielding thermal diffusivities that agreed with literature values within 2%. We then measured the thermal diffusivities in the nucleus and cytoplasm of living COS7 cells and obtained 0.134 (±0.008) $\times 10^{-6}$ $m^2s^{-1}$ and 0.133 (±0.002) $\times 10^{-6}$ $m^2s^{-1}$, respectively. These values correspond to 94% (±1.7%) and 93% (±5.6%) of that of water. These results demonstrate that intracellular heat transfer is close to that of an aqueous environment and do not support the hypothesis that substantially reduced intracellular heat conduction underlies the $10^5$-gap discrepancy.

We next directly compared heat-induced temperature changes in living cells by label-free MIP-ODT thermometry and fluorescent nanothermometry. Continuous heating of a selected intracellular region for several seconds produced a step-like temperature response in the MIP-ODT readout, with rapid rise and decay. This temporal behavior is consistent with the water-like intracellular thermal diffusivity quantified above. Under identical heating conditions, the fluorescent probe exhibited the same initial step-like response but additionally showed a slowly varying component. This gradual signal was absent in the MIP-ODT readout, indicating that it does not represent LTE-defined temperature.

We then investigated the origin of this slowly varying component. Control experiments showed that it is unlikely to arise from methodological artifacts of fluorescent nanothermometry and instead reflects an intrinsic intracellular phenomenon. We further found that the signal is detected through the molecular transition involved in LTE-defined temperature sensing by fluorescent nanothermometers. Moreover, the signal is unlikely to originate from processes directly coupled to major metabolic pathways and appears to require intact cellular organization and biomolecular assemblies. Together with its emergence on timescales of seconds, these observations suggest that the slowly varying component is associated with intracellular processes whose relaxation dynamics are slower than conductive heat dissipation. One possible explanation is transient energy storage in conformational or structural rearrangements of biomolecules. This view is also consistent with the difference in measurement principle between fluorescent nanothermometry and MIP-ODT, because fluorescent nanothermometers, unlike MIP-ODT, can be sensitive to such local non-equilibrium energy states.

Identifying a non-conductive intracellular component provides a framework for interpreting intracellular thermometry and the $10^5$ gap issue. Our results show that intracellular heating generates two distinct responses: an LTE-defined temperature rise and a long-lived non-conductive component. Fluorescent nanothermometry is sensitive to both contributions and therefore cannot selectively report conductive thermal dynamics. From this perspective, the $10^5$-gap discrepancy may arise from comparing an LTE-defined temperature with a slowly evolving non-conductive intracellular signal. These findings clarify the interpretation of fluorescence-based temperature readouts and open a route to probing previously unrecognized intracellular energy dynamics.

## Methods

### MIP-ODT system.

A detailed schematic of the MIP-ODT system is represented in Fig. S1 of Supplementary Note 1. A homemade Ti-Sapphire laser at a repetition rate of 1 kHz serves 10-ns visible probe pulses at 705 nm, which are guided into a Mach-Zehnder interferometer through an optical single-mode fiber. In the sample arm of the interferometer, the probe pulses go through a beam steering optics consisting of a grating with lines per millimeter of 70 (46-068, Edmund), an aperture, relay lenses, and a reflective objective lens (5007-000, Beck Optronic Solutions) placed in a 4-f configuration so that the beam is illuminated onto a sample at a tilted angle with a numerical aperture (NA) of 0.58. The illumination angle can be discretely changed by rotating the grating at ten different angles. The aperture is synchronously rotated with the grating to transmit only the first diffraction beam. The FWHM illumination area on the sample is 110 μm in diameter, which is a 1/11.8 times magnification from that on the grating. The sample image is magnified in the image sensor plane for holographic measurement by a factor of 167 with an objective lens (LCPLFLN100XLCD, Olympus) and relay lenses. The reference light is directed to the image sensor (Q-2HFW, Adimec) in the off-axis configuration after adjusting the optical path length, beam size, and polarization to match those of the object light.

We used two IR light sources, a homemade ns-PPLN-OPO[36] (ns-IR) and a 1,456-nm CW laser diode (LD) (BL1456-PAG500, Thorlabs) (CW-IR), for the impulsive and continuous heating, respectively. The ns-IR laser is operated at a repetition rate of 1 kHz, and the output wavenumber is tunable from 2,800 to 3,250 cm$^{-1}$. The ns-IR beam is mechanically modulated with an optical chopper, while the CW-IR beam is electronically controlled with an LD driver. For impulsive and continuous dual-heating, the ns-IR and CW-IR beams are spatially combined with a dichroic mirror (106950, LAYERTEC) and subsequently combined with the visible beam with another dichroic mirror (100-00-024, S.T.Japan). These beams are focused onto the sample through the reflective objective lens.

**Selection of IR wavenumber.**

To ensure intracellular local heating, we carefully selected the optimal IR wavenumber with which the IR light is absorbed inside the cell. Figure S2a in Supplementary Note 2 illustrates the depth (z-axis) profiles of $\delta n_\rho$ for water, which were acquired through MIP-ODT measurements with ns-IR wavenumbers of 3,000, 3,050, and 3,150 cm$^{-1}$. Solid lines in the figure show experimental results, which are in good agreement with the dashed lines derived from the theoretical absorption profiles assessed by molar extinction coefficients at the corresponding IR wavenumbers, showing the 1/e decays of 7, 5, and 2 μm, respectively[37]. We also confirmed that the cell and culture medium absorb a similar amount of IR light at 3,150 cm$^{-1}$ (see Fig. S2b). Hence, we employed 3,150 cm$^{-1}$ for IR wavenumber to ensure intracellular local heating, where most of the light (~90%) is absorbed by the adherent cells with a typical thickness of ∼5 μm (Fig. S2c). For the thermal diffusivity measurement presented in Fig. 2, between 77 to 99% of the IR light was absorbed by the cells, depending on the cellular thickness. In the experiment for comparison between label-free and fluorescence nanothermometry (shown in Fig. 3), we selected 3,100 cm$^{-1}$ with 1/e attenuation of around 3 μm, to obtain a signal from the entire cell in depth for a better comparison with fluorescence measurements because fluorescence imaging provides depth-integrated signals. The wavelength of CW-IR is 1,456 nm, which exhibits a lower molar extinction coefficient for water. Although it does not ensure intracellular local heating along the z-axis, the z-sectioning capability of ODT enables measuring intracellular signals.

**Thermal diffusivity measurement.**

Figure S3 in Supplementary Note 3 depicts the timing chart and control system of MIP-ODT for pump-probe imaging with nanosecond pulses. Repetition rate and timing of the ns-IR pulses (1 kHz), visible pulses (1 kHz), image sensor (100 Hz), and optical chopper (50 Hz) were electrically controlled with a function generator (DG1022Z, RIGOL), and the sensor alternately acquired IR-ON and -OFF frames (see Fig. S3a). The sensor recorded ten visible pulses in a single frame. The delay and illumination angle were also controlled by the function generator (see Fig. S3b).

We acquired 200-frames-averaged phase images by digital holography at ten different angles of illumination for each pump-probe delay and reconstructed an RI image. For RI reconstruction, non-negative constraint regularization was applied with 100 iterations. We selected an initial delay for transient measurement at 3 μs because heat conduction towards the glass dominantly occurs within a few μs after the heating. To precisely evaluate the heat-induced $\delta n_{\rho}$, we integrated the RI images stacked in the z-direction and visualized the two-dimensional flow of thermal energy.

For intracellular measurements, we carefully confirmed that the heated volume, particularly in the z direction, is within the cell by depth-sectioned ODT images. A slight heat distribution into the surrounding extracellular water (~10% for a typical cellular thickness of 5 μm) was confirmed to cause only a minor deviation of a few percent in the measured thermal diffusivity (see Supplementary Note 4 for details). Additionally, as demonstrated in Supplementary Note 5, we experimentally confirmed that the measured thermal diffusivity remains consistent across varying levels of intracellular IR absorption, ranging from 77% to 99%, depending on the cellular thickness. This results also verifies that the slight heating of extracellular water does not affect our intracellular heat conduction measurements.

The residual error of the thermal diffusivity measurement cannot be explained by the noise of our visible ODT measurement, and we suspect that it comes from the IR irradiation conditions, such as fluctuations of the laser power or the beam spot size, which could be reduced by monitoring the laser power or implementing a more robust optical system.

**Calibration curve of $\boldsymbol{\delta n_{\rho}}$ on $\boldsymbol{\delta T}$.**

We performed calibration of $\delta n_{\rho}$ of water with respect to the temperature rise $\delta T$ because it is known that there is a slight temperature dependence in the thermo-optic coefficient (dn/dT). It should be noted that the nonlinearity derived below is a minor effect that causes deviation in thermal diffusivity by 4% than that without the calibration. Details are described in Supplementary Note 6.

**Heat conduction calculation.**

The heat conduction equation is written as,

$$\frac{\partial \delta T(x,y,z,t)}{\partial t} = \left(\frac{\partial^2}{\partial x^2} + \frac{\partial^2}{\partial y^2} + \frac{\partial^2}{\partial z^2}\right) \alpha \delta T(x,y,z,t), \quad \text{(Eq. 4)}$$

where $\delta T$ is the temperature change, $x$, $y$, and $z$ are the spatial coordinates, $t$ is the time, and $\alpha$ is the thermal diffusivity. We used the experimentally obtained $\delta T$ map as an initial profile in the x-y plane and a theoretical absorption profile along the z-axis. Then, we calculated the evolution of $\delta T$ map by imposing the glass-sample boundary condition. For rigorous comparisons with the experimental results, the instrumental function of ODT was applied to the calculation results. Since the dn/dT of glass is significantly smaller than those of water and glycerol[38], $\delta n_{\rho}$ in the glass area can be considered to be zero in the calculation, corresponding to the negligibly small $\delta n_{\rho}$ experimentally observed in ODT measurement even if heat transfers into the glass.

We used a self-made program based on the Forward Time Centered Space (FTCS) method, which was implemented using Python. The pixel pitch was set to 207 nm to match the ODT measurement. When the sample was water, the IR absorption profiles along the z-axis, required for generating the initial temperature map, were determined based on the literature value of the absorption coefficient for water[37]. For glycerol and cells, we could not find literature values. Therefore, we used a 1/e decay length of 3.5 μm for glycerol, derived from the experimental data, and the same value as water for cells because the absorption profile shows good agreement with that of water, as demonstrated in Fig. S4.

We imposed the boundary conditions between the sample and the glass substrate written by

$$K_{\text{sample}}\left(\frac{\partial T_{\text{sample}}(x,y,z,t)}{\partial z}\right)_{z=\text{boundary}} = K_{\text{glass}}\left(\frac{\partial T_{\text{glass}}(x,y,z,t)}{\partial z}\right)_{z=\text{boundary}}, \quad \text{(Eq. 5)}$$

where $K_{\text{sample (glass)}}$ and $T_{\text{sample (glass)}}$ are the thermal conductivity and temperature of the sample (glass), respectively. The thermal conductivity of the glass substrate was set to 0.76 $Wm^{-1}K^{-1}$ for soda-lime glass, specified by the manufacturer, from which we derived the thermal diffusivity of $0.40\times10^{-6}$ $m^2s^{-1}$, and those of water and glycerol are 0.61 and 0.29 $Wm^{-1}K^{-1}$, respectively[39,40]. The thermal conductivity of cells was calculated by multiplying the thermal diffusivity with the specific heat capacity of 3,900 J/(kg·K) and the density of 1.08 $g/cm^3$ reported in previous studies[24,25].

To apply the instrumental function of the ODT measurement to a 3D temperature map simulated by the heat conduction calculation, we converted the temperature map to an RI map and calculated phase images through the light propagation equation for all the illumination angles by the angular spectrum method. Then, these phase images were used as inputs for the ODT reconstruction algorithm to produce a 3D RI image with the instrumental function of the ODT measurement.

**Sample preparations.**

COS7 cells were cultured either on a soda-lime glass plate (1-9648-01, AS ONE) (for experiment shown in Fig. 2) or a $CaF_2$ plate (C20SQ-0.5, Pier Optics) (for experiment shown in Fig. 3) with high-glucose Dulbecco's modified eagle medium (DMEM), which contains L-glutamine, phenol red, and HEPES (FUJIFILM Wako). The medium was supplemented with 10% fetal bovine serum (Cosmo Bio) and 1% penicillin-streptomycin-L-glutamine solution (FUJIFILM Wako) at 37°C in a 5% $CO_2$ atmosphere. The COS7 cells were sandwiched between the substrates for imaging.

We performed microinjection to introduce FPTs into cells following the subsequent procedure: (1) FPTs were loaded into a capillary (524-295-7000, Eppendorf) equipped with a chip (5242-956-003, Eppendorf). The capillary was affixed to the injector (FemtJet, Eppendorf). (2) While monitoring the microscopic field-of-view (FOV), the tip of the capillary was gradually lowered to the surface of the dish using a controller (Micromanipulator, Eppendorf). (3) Subsequently, FPTs were injected through the cellular membrane while maintaining the pressure in the capillary within the range of 50 to 100 hPa. The injection solution comprised FPTs dissolved in water at a concentration of 0.5 weight/volume (w/v)%, whereas the estimated concentration of FPTs within the cell was approximately 0.02 w/v%. The environmental temperatures during measurements were 24°C for thermal diffusivity measurements and 26°C for heat-induced intracellular temperature change measurements.

**Fluorescence imaging.**

Our MIP-ODT system incorporates an epifluorescence imaging modality (see Fig. S1 in Supplementary Note 1). We used FPT for ratiometric sensing, in which fluorescence wavelengths were 500-520 nm and 550-600 nm (FDV-0005, FNA) under 488-nm excitation (WSLP-488-010m-4-B, Wavespectrum). The detection wavelength band was selected by optical bandpass filters (FBH520-40 and FELH0550, Thorlabs). The average excitation power was ~40 μW with an illumination spot diameter of ~50 μm. The fluorescence was then transmitted through a dichroic mirror (MD499, Thorlabs) and projected onto an sCMOS image sensor (Zyla5.5, Andor). The image sensor was operated at 25 fps. The time variation of temperature was evaluated by calculating the ratio of the fluorescence intensities at the two wavelength bands. The detailed procedure is outlined in Supplementary Note 8.

**Heat-induced intracellular temperature change measurement.**

Figure S9 in Supplementary Note 9 shows the timing chart for the continuous heating experiment. The CW-IR beam was repetitively illuminated onto the sample by heating and cooling cycles of 2.4 s and 12.6 s, respectively, in addition to the impulsive heating with the ns-IR pulses (Fig. S9a). Figure S9b shows the timing chart for fluorescence measurements. The excitation light for fluorescence imaging was illuminated onto the sample for a duration of 4.6 s, starting before the initiation of continuous heating by 0.8 s.

In the experiment corresponding to Figs. 3 and S11, the average power of CW-IR light was ~2 mW. The result of time variation displayed in Fig. S11d was obtained without averaging, whereas that in Fig. 3b was calculated as the average of two measurements. The $\Delta n$ values in these plots were calculated by averaging 2.1×2.1×4.2 μm for Fig. S11d and 2.8×2.8×5.6 μm for Fig. 3b along x-, y-, and z-axes in the center of the continuous heating spot.

**Dual-heating method.**

In the dual-heating method, $\Delta n$ with impulsive heating OFF and ON are expressed as

$$\Delta n_{\mathrm{IH-OFF}} = \Delta n_\rho + \Delta n_\sigma, \quad \text{(Eq. 6)}$$

$$\Delta n_{\mathrm{IH-ON}} = \Delta n_\rho + \Delta n_\sigma + \delta n_\rho = \Delta n_\rho + \Delta n_\sigma + \left(\frac{dn}{dT}\right)_{T=T_0+\Delta T} \delta T, \quad \text{(Eq. 7)}$$

where $T_0$ denotes the initial temperature right before the continuous heating, and $\Delta T$ and $\delta T$ are the temperature changes induced by continuous heating and impulsive heating, respectively. The terms $\Delta n_\rho$ and $\Delta n_\sigma$ represent RI changes induced by continuous heating. Specifically, $\Delta n_\rho$ arises from temperature-dependent density changes associated with thermal expansion, whereas $\Delta n_\sigma$ originates from slower heat-induced molecular redistribution. The term $\delta n_\rho$ represents the RI change induced by the impulsive heating. Note that the symbols $\Delta$ and $\delta$ on the right-hand side of the equations denote changes induced by continuous and impulsive heating, respectively. The term $\delta n_\rho$ depends on $\Delta T$ because the RI change measurement with impulsive heating was operated at every 20 ms, where the initial temperature for each impulsive heating was $T_0 + \Delta T$. Selective detection of $\delta n_\rho$ is achieved by subtracting the impulsive heating OFF state from the ON state, thereby canceling $\Delta n_\rho$ and $\Delta n_\sigma$. For accurate isolation of $\delta n_\rho$, the differential measurement must be performed within a time window sufficiently short to avoid contamination from slowly varying effects. To satisfy this condition, we used a frame interpolation method (see Supplementary Note 13 for details). Using this approach, the residual slowly varying contribution from $\Delta n_\rho$ and/or $\Delta n_\sigma$ was reduced to less than 1% of total $\Delta n$. We further corrected the slowly varying minor effect (see Supplementary Note 14 for details), although the characteristics remain unchanged without the correction. The IR-visible delay was set to 1 μs to maximize the SNR. The differential measurement between impulsive heating ON and OFF yields Equation 2. Since $\delta T$ and $T_0$ are constant, the temperature change due to continuous heating, $\Delta T$, can be determined from the temperature dependence of the thermo-optic coefficient.

**Calibration curve of $\delta n_\rho$ on $\Delta T$.**

To accurately determine $\Delta T$ with the dual-heating method, we measured $\Delta T$ dependence of $\delta n_\rho$ by changing the amount of continuous heating. The absolute value of $\Delta T$ was determined by $\Delta n_\rho$ divided by intracellular thermo-optic coefficient dn/dT (see Supplementary Note 15 for details). We evaluated $\Delta n_\rho$ by selectively measuring $\Delta n$ at the very beginning of the continuous heating, 5 ms after the beginning of the heating. We applied the non-negative constraint to evaluate $\Delta n_\rho$ for better accuracy because accurate determination of $\Delta T$ was required. For an accurate evaluation, we averaged 200 images of $\Delta n$. On the contrary, we did not apply the non-negative constraint to evaluate $\delta n_\rho$ because the primary importance lied in precise determination of relative values rather than accuracy.

From the abovementioned measurement, we found that $\delta n_\rho$ linearly increases against $\Delta T$ as

$$\left(\delta n_\rho\right)_{T=T_0+\Delta T} / \left(\delta n_\rho\right)_{T=T_0} \approx 1 + a\Delta T, \qquad \text{(Eq. 8)}$$

where $a$ is a constant value. Figure S15 in Supplementary Note 16 shows the measured data from which we evaluated $a$. Note that each cell provides slightly different $a$ ($9.8 \times 10^{-3} \pm 8.9\%$). This is probably because molecular composition depends on the measurement area. Therefore, we evaluated $a$ specifically for the measured cell to determine $\Delta T$ accurately.

**Investigation of potential artifacts in fluorescent nanothermometry.**

Microinjection of FPT, CP [ref 32, see Fig. E in Supplementary information], and Rhodamine B-labeled dextran (molecular weight 10,000; Thermo Fisher Scientific) into living cells was performed using Femtojet microinjection system (Eppendorf, Hamburg, Germany), controlled by a micromanipulator (Eppendorf). FPT (0.5 w/v%), CP (1.0 w/v%), and Rhodamine B-labeled dextran (20 µM) were dissolved in an aqueous buffer containing 140 mM KCl, 10 mM $KH_2PO4$–$K_2HPO_4$, and 4 mM NaCl. The solution was delivered into the cytoplasm using a glass capillary needle (Femtotips II; Eppendorf). The injected volume was estimated to be ~2 fL.

Liposome-encapsulated water containing Rhodamine B-labeled dextran was prepared as follow. 1,2-Dioleoyl-sn-glycero-3-phosphocholine (DOPC) was dissolved in chloroform/methanol mixtures (2:1, v/v). An aliquot of 20 µL of DOPC solution (5 mM) was transferred into a glass test tube and gently dried under a stream of air to form a thin DOPC film on the bottom of the tube. This film was subsequently dried under a vacuum for 2 h. Hydration was then performed using 190 µL of Rhodamine B-labeled dextran (10 µM) in a 200 mM sucrose solution at 50 °C overnight, following a 5 min prehydration step at 50 °C using 10 µL of the same solution. The final DOPC concentration was 0.5 mM.

To suppress liposome displacement caused by convection during IR laser irradiation, a microscale chamber array containing wells of 20 µm diameter and 50 µm depth was employed. Individual liposomes were positioned within the wells. A coverslip containing SU-8 fabricated microchambers was first treated with 200 µL of 200 mM glucose solution and subjected to deaeration. Subsequently, 4 µL of the liposome suspension and 36 µL of 200 mM glucose solution were added. The coverslip was then sealed with a cover glass to enable fluorescence imaging.

Fluorescent nanothermometry was performed using confocal fluorescence imaging on a TCS SP8-FALCON confocal laser scanning microscope (Leica Microsystems). Excitation wavelengths of 552 nm for Rhodamine B and 488 nm for FPT and CP were used. Fluorescence emission was collected at 560-700 nm for Rhodamine B, 500-520 nm and 550-600 nm for FPT, and 550-600 nm for CP. IR heating was applied using a 1480 nm laser source (FRL-DC, IRE-Polus, MA, USA). The diameter of the focused infrared irradiation area was estimated to be 1.29 µm with a laser power of 0.83 mW. The temperature of the culture medium was maintained at 29 °C during measurements. Temperature regulation and monitoring were performed using a stage-top incubator and an objective lens heater (INUBSF-ZILCS; Tokai Hit, Fujinomiya, Japan). Because Rhodamine B does not operate as a ratiometric thermometer, thermophoretic motion of the probe can influence the fluorescence readout. To mitigate this effect, intracellular temperature changes were determined from the spatially averaged fluorescence signal over the entire cell. All analyses presented in Fig. 4 were therefore performed using whole-cell averaged fluorescence signals.

**Data availability**

The data provided in the manuscript are available from the corresponding author upon reasonable request.

**Acknowledgments**

This work was financially supported by Japan Society for the Promotion of Science (20H00125, 23H00273), JST PRESTO (JPMJPR17G2), JST FOREST Program (JPMJFR236C), Precise Measurement Technology Promotion

Foundation, Research Foundation for Opto-Science and Technology, Nakatani Foundation, and UTEC-Utokyo FSI Research Grant.

## Author contributions

K.T. developed the MIP-ODT system, performed the experiments, and analyzed the experimental data. K.T. and M.T. performed the experiments for the investigation of potential artifacts in fluorescent nanothermometry. M.T. and K.O. provided the FPT-injected COS7 cells. K.T., M.T., H.S., K.O., and T.I. discussed the interpretation of the results. V.R.B. constructed the Ti-Sapphire laser and OPO. H.S. wrote codes for data acquisition. K.T. and G.I. wrote codes for heat conduction simulation. T.I. supervised the entire work. K.T. and T.I. wrote the manuscript with inputs from the other authors.

## Competing interests

K.T. and T.I. are inventors of patents related to the MIP-ODT system.

# Supplementary Information for
# Label-free mid-infrared photothermal microscopy revisits intracellular thermal dynamics: what do fluorescent nanothermometers measure?

Keiichiro Toda[1,2], Masaharu Takarada[3], Genki Ishigane[2], Hiroyuki Shimada[1], Venkata Ramaiah Badarla[1], Kohki Okabe[3] and Takuro Ideguchi[1,2,4,5,*]

[1] Institute for Photon Science and Technology, The University of Tokyo, Tokyo, Japan

[2] Department of Physics, The University of Tokyo, Tokyo, Japan

[3] Department of Pharmaceutical Science, The University of Tokyo, Tokyo, Japan

[4] Department of Advanced Materials Science, The University of Tokyo, Tokyo, Japan

[5] RIKEN Center for Advanced Photonics, RIKEN, Saitama, Japan

[*] Corresponding author: ideguchi@edu.k.u-tokyo.ac.jp

**Supplementary Note 1: Detailed schematic of MIP-ODT system.**

Figure S1 represents a detailed schematic of our MIP-ODT system, which also integrates epifluorescence microscopy.

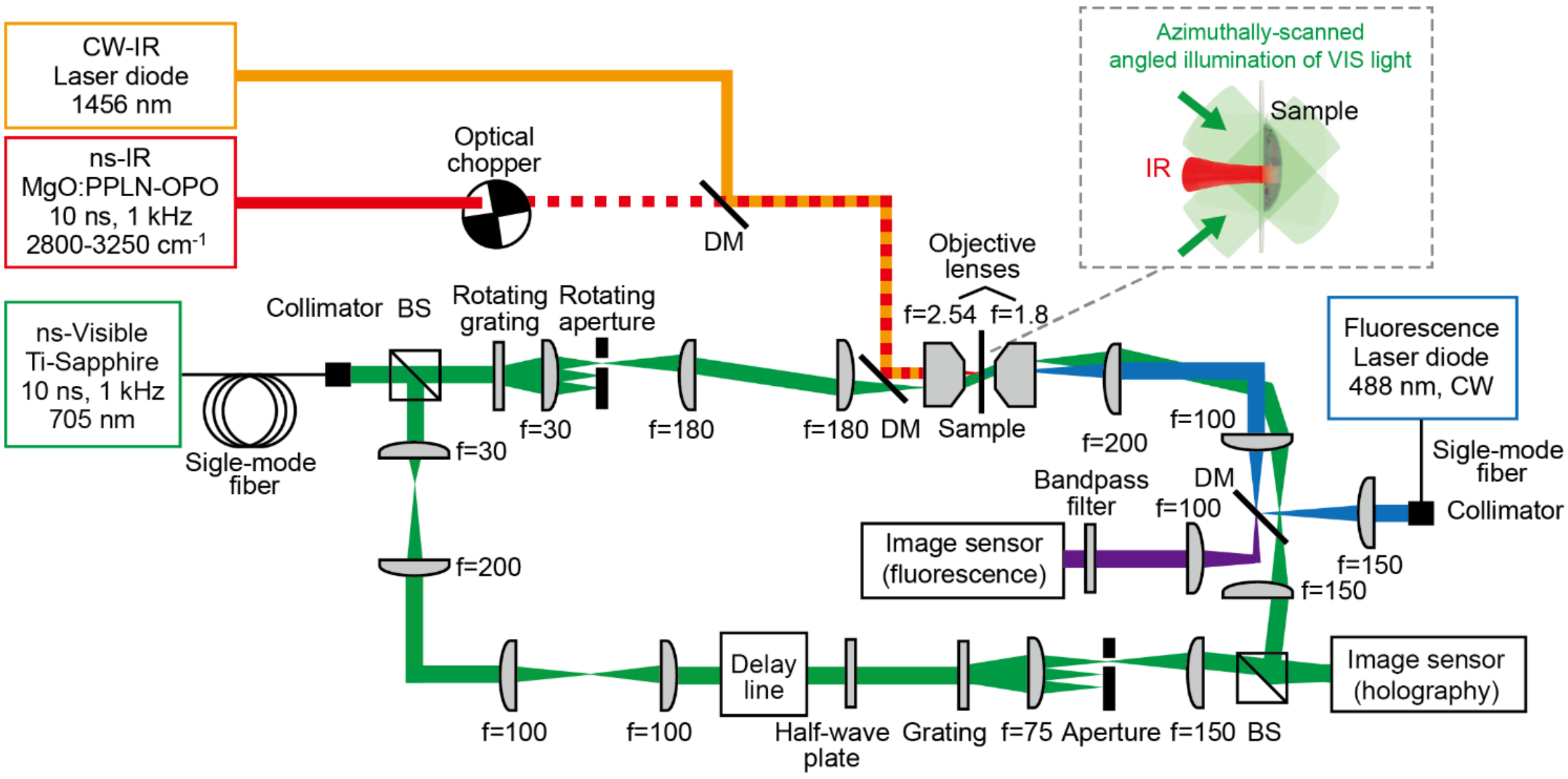

**Fig. S1 Detailed schematic of MIP-ODT system.** PPLN-OPO: Periodically-poled lithium niobate optical parametric oscillator, DM: Dichroic mirror, BS: Beamsplitter.

**Supplementary Note 2: IR wavenumber selection for intracellular thermal diffusivity measurement.**

Figure S2 shows depth (z-axis) profiles of the $\delta n_{\rho}$ measured by MIP-ODT with ns-IR wavenumbers of 3,000, 3,050, and 3,150 cm$^{-1}$.

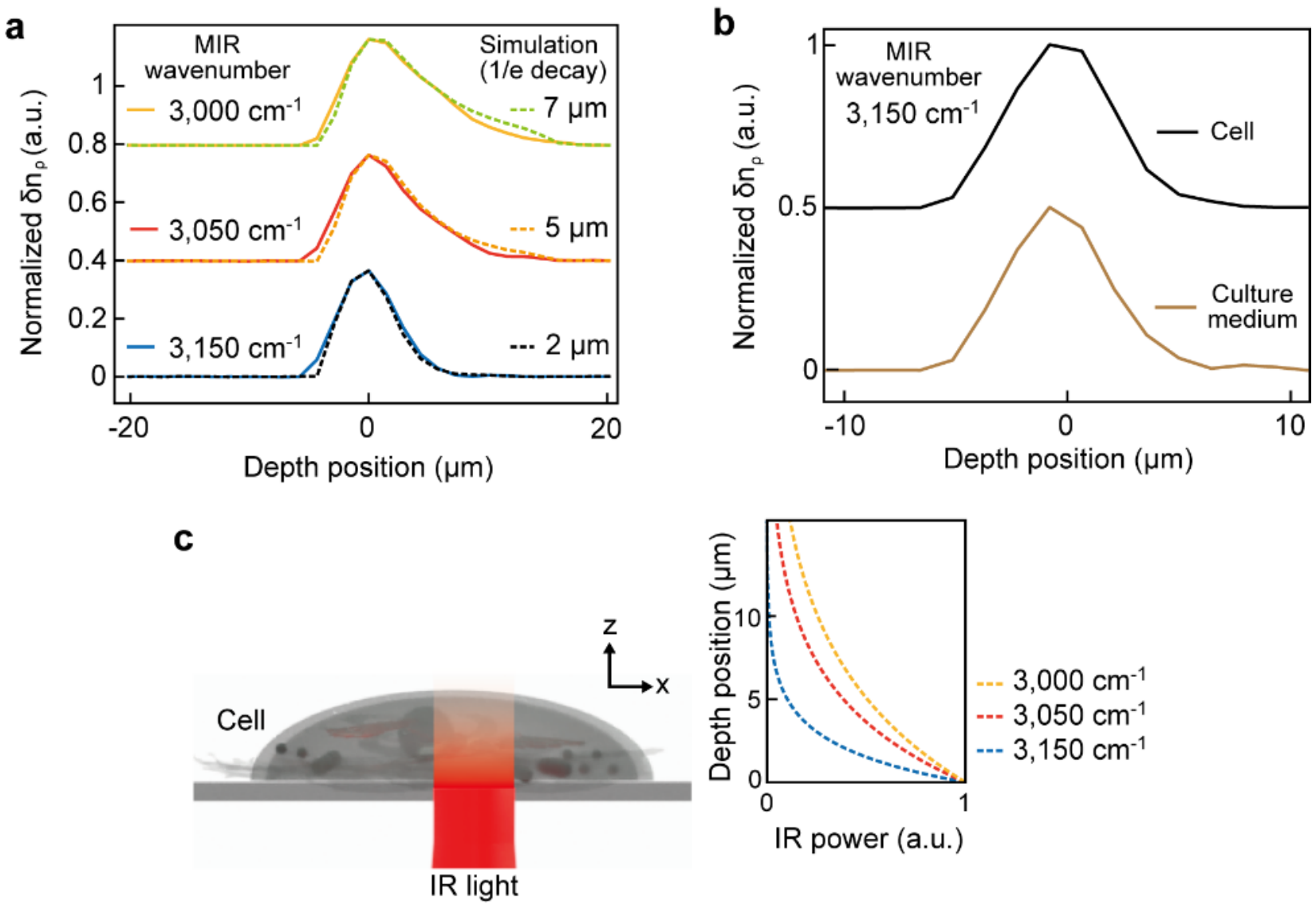

**Fig. S2 Measured and simulated attenuation depth of the IR light. a** Solid lines: Depth (z-axis) profiles of normalized $\delta n_p$ for a water sample measured by MIP-ODT excited at IR wavenumbers of 3,000 (yellow), 3,050 (red), and 3,150 $cm^{-1}$ (blue). Dashed lines: Simulated depth profiles by convolving the instrumental function of our ODT measurement to single exponential functions with 1/e decay lengths of 7 (light green), 5 (orange), and 2 µm (black). **b** Depth profiles of normalized $\delta n_p$ for a culture medium (brown) and a COS7 cell (black). **c** Illustrative depiction of attenuation profiles of IR light illuminating a cell with a thickness of 5 µm at different IR wavenumbers.

**Supplementary Note 3: Timing chart and control system of MIP-ODT for pump-probe imaging.**

Figure S3 displays the timing chart and the control system of MIP-ODT for pump-probe thermal diffusivity measurements.

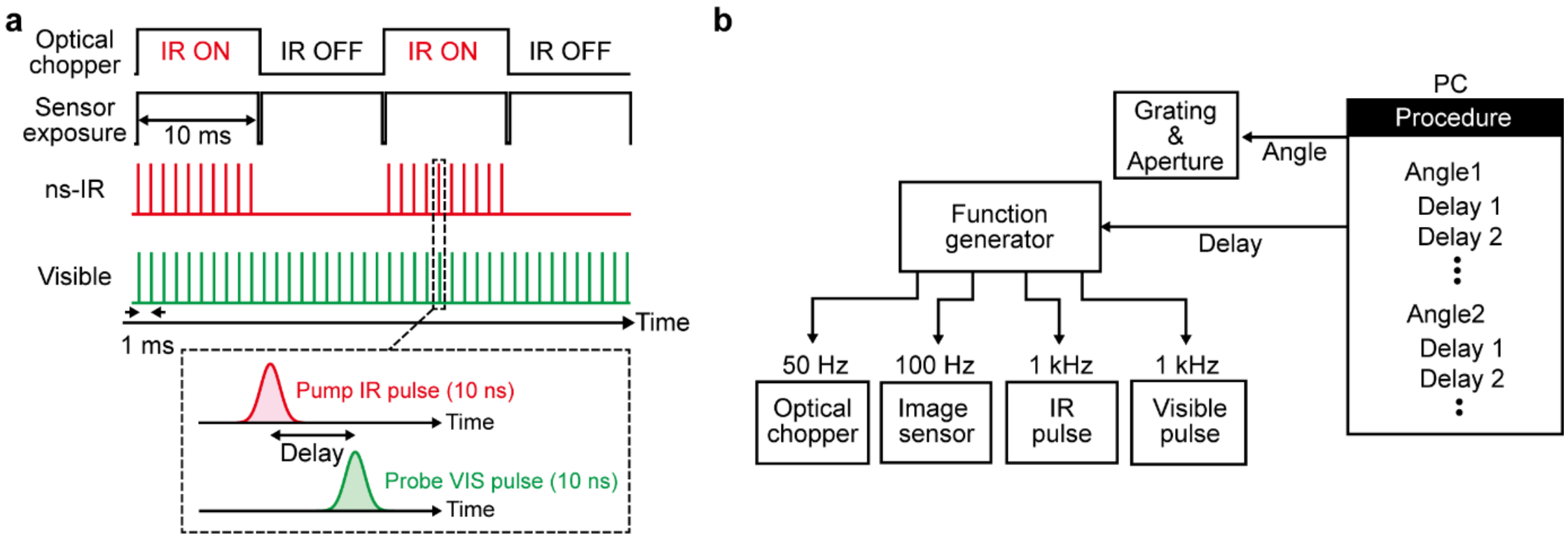

**Fig. S3 Timing chart and control system of MIP-ODT for pump-probe imaging with nanosecond pulses. a** Timing chart of the optical chopper, sensor exposure, and ns-IR/visible pulses. **b** Control system. A function generator controls the frequency and phase (timing) of the optical chopper, image sensor, IR pulses, and visible pulses. A PC sets the illumination angle and pump-probe delay.

**Supplementary Note 4: Numerical investigation of influence of slight heat distribution in the surrounding water on the thermal diffusivity measurement.**

To investigate the impact of a small amount of heat distribution in the water behind the cell on the thermal diffusivity measurements, we compared simulations and experimental results of the z-axis profiles of $\delta n_\rho$ for water and a cell on a glass substrate. Figures S4a and S4b display time-resolved profiles of $\delta n_\rho$ for water obtained through heat conduction simulation with and without applying the instrumental function of the ODT measurements. Figures. S4c and S4d show experimental results for water and a cell measured with MIP-ODT. We verify that the experimental results show good agreement with the simulation, demonstrating our heat conduction calculations, depicted in Fig. S4a, accurately reproduce the actual thermal diffusion. It is also clearly observed that the experimentally measured profile of water (Fig. S4c) is nearly identical to that of a cell (Fig. S4b). These comparisons guarantee that we can use the simulation result shown in Fig. S4a to evaluate the cell measurements. The ratio of the $\delta n_\rho$ outside the 5-µm sample region against the total $\delta n_\rho$ within the integrated volume is ~8.9% at a delay of 0 s, which gradually increases via thermal diffusion (Fig. S4a). The mean value becomes ~13% during the delay shown in Fig. 2b (3-13 µs), which indicates that 87% of the induced heat resides in the 5-µm-thick cell, while 13% of that presents in the surrounding water. This amount of heat leakage does not majorly alter the value of intracellular thermal diffusivity. In our specific case, a simple estimation via a weighed linear summation of the two components, $0.133\times10^{-6}$ $m^2s^{-1}\times0.87$ (cell) + $0.143\times10^{-6}$ $m^2s^{-1}\times0.13$ (water), provides a thermal diffusivity of $0.134\times10^{-6}$ $m^2s^{-1}$, which is deviated by 1% from that without the correction.

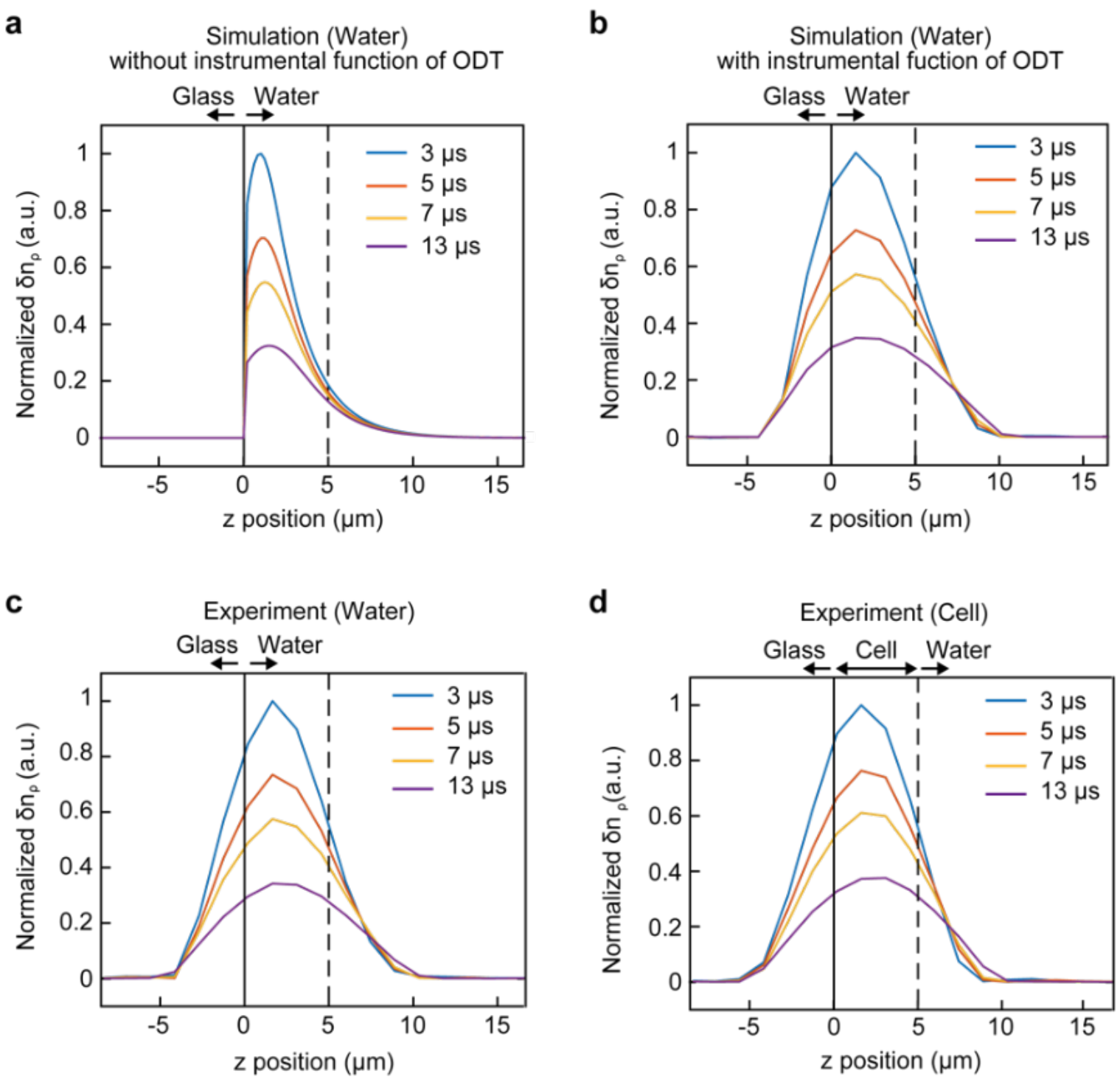

**Fig. S4 Time-resolved z-axis profiles of $\delta n_\rho$ for water and a cell. a** Simulation results for water without the instrumental function of ODT. The 1/e attenuation of IR light is set to 2 µm, which is the same condition as the experiment shown in Fig. 2. **b** Simulation results for water with the instrumental function of ODT. **c** Experimental results for water. **d** Experimental results for cell.

**Supplementary Note 5: Experimental investigation of influence of slight heat distribution in the surrounding water on the thermal diffusivity measurement.**

To experimentally validate the impact of a small amount of heat distribution in the water behind the cell on the thermal diffusivity, we evaluate whether cell thickness affects the measured thermal diffusivity. This is because the amount of IR absorption in the extracellular water depends on cell thickness. In cells with various thicknesses, the distance between the glass substrate and the surrounding water differs, consequently altering the proportion of IR absorption inside and outside the cells. To evaluate cell thickness, we measured the optical phase delay induced by cells, which is known to exhibit a linear relationship with cellular thickness. Figure S5 illustrates the relationship between phase delay (cell thickness) and measured thermal diffusivity, demonstrating that the measured thermal diffusivity remains consistent even for cells with significantly varying thicknesses (with corresponding IR absorption ranging from 77% to 99%).

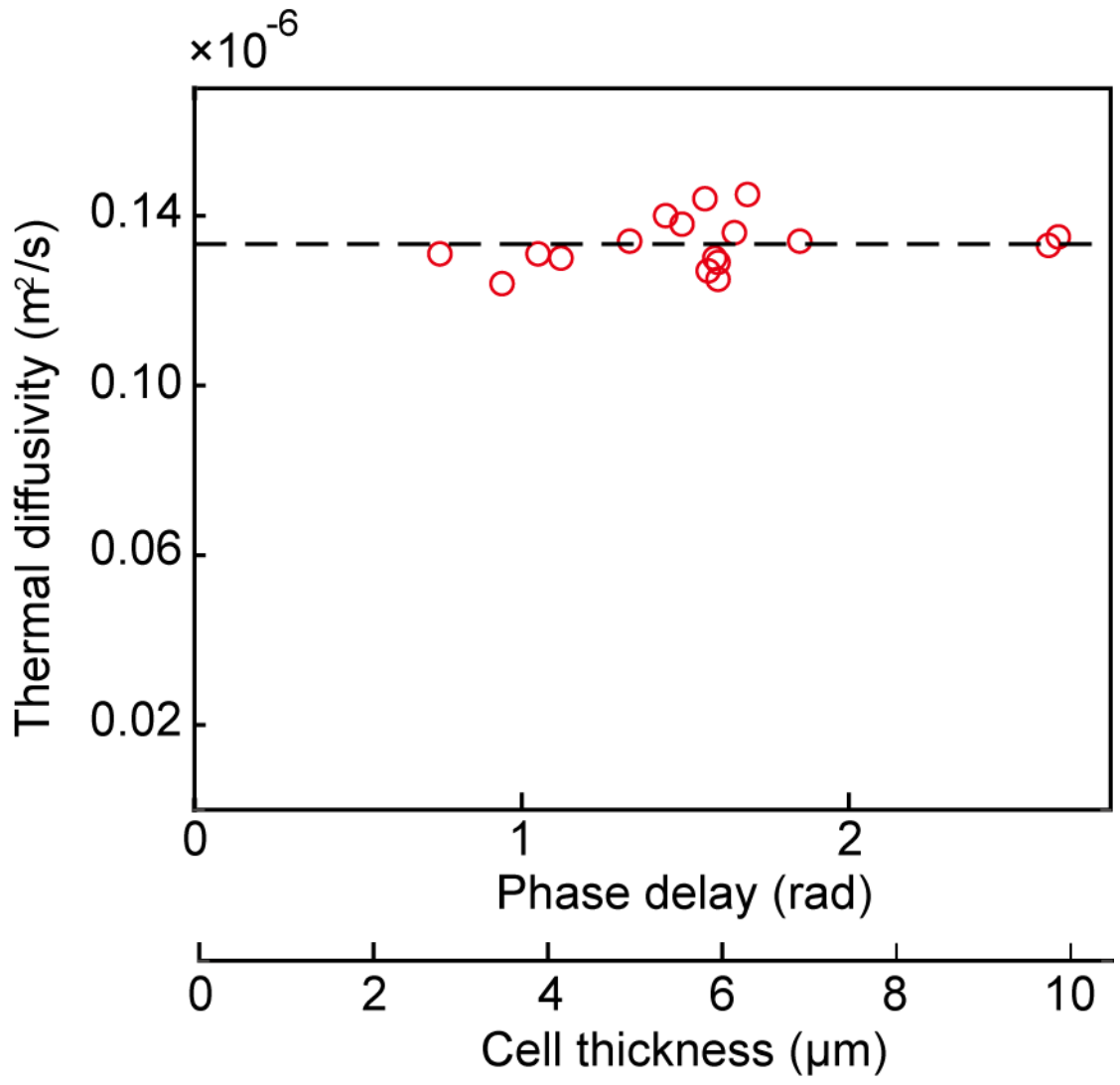

**Fig. S5 Measured thermal diffusivity for various cellular thickness.** The cellular thickness was estimated from the phase delay measured by ODT. For this calculation, we assumed an RI difference of ~0.03 between the cell and the culture medium, derived from the RI increment of proteins (0.19 l/mg)[1] and the typical dry mass concentration within a cell (~0.15 g/ml)[2]. The black dashed line represents the average intracellular thermal diffusivity measured in this study.

**Supplementary Note 6: Calibration curve of $\delta n_\rho$ on $\delta T$.**

Considering the numbers of IR photons and water molecules in the excited volume with their vibrational decay time, the IR pulse used in our study can excite a small number of water molecules in the excited volume. Therefore, the vibrational absorption of water in this work is assumed to be in the linear regime without saturation. Since the temperature rise can be expected to be in a linear relation with the amount of absorption, we investigate the dependence of $\delta n_\rho$ on the IR power to make a calibration curve.

To evaluate the dependence of $\delta n_\rho$ on the IR power, we measured crossectional $\delta n_\rho$ profiles with different IR powers (Fig. S6a). We first use a linear relationship between $\delta n_\rho$ and $\delta T$ with an equation of $\delta n_\rho = 10^{-4}\delta T$ (Fig. S6b), which is a known value of the thermo-optic coefficient of water (dn/dT = $1 \times 10^{-4}$) at room temperature. Figure S6c displays $\delta T$ profiles normalized by the excitation IR power 1, showing nonlinearity, particularly around the center of the heated spot, which owes to the slight temperature dependence of dn/dT. According to the abovementioned linear relationship between the temperature rise and the IR power, the normalized $\delta T$ profiles by excitation IR power must be identical. Therefore, we assume a nonlinear relationship with a quadratic term as $\delta n_\rho = 10^{-4}(a\delta T + b\delta T^2)$ and find the constants (a and b) to make the normalized profiles overlap (Fig. S6d). We obtain a=0.98 and b=0.02, respectively, which agree well with the literature values[3]. The nonlinear calibration curve is shown in Fig. S6b.

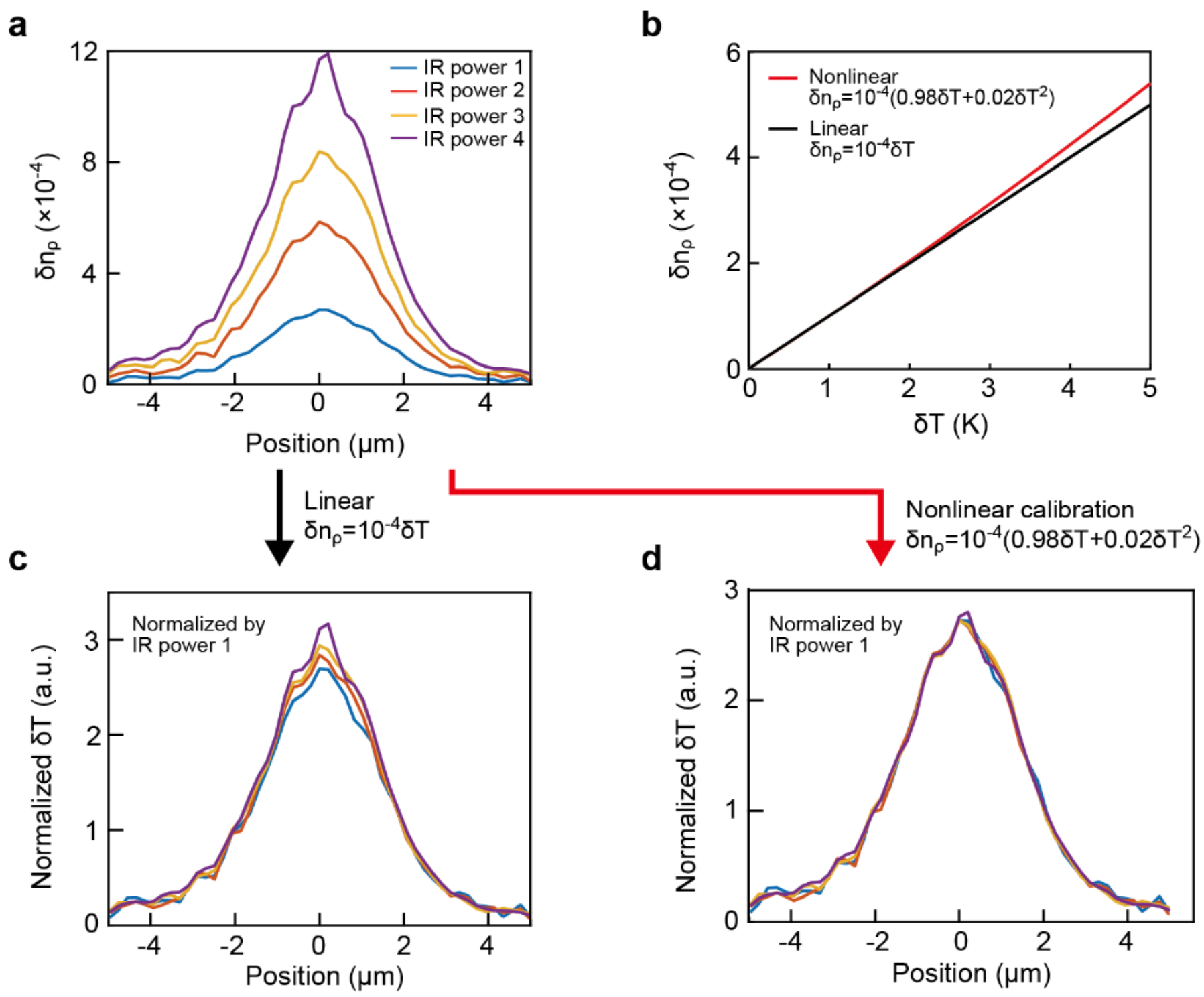

**Fig. S6 Determination of the calibration curve of $\delta n_\rho$ on $\delta T$. a** Crossectional profiles of $\delta n_\rho$ of water measured with different IR powers. **b** The nonlinear calibration curve of $\delta n_\rho$ on $\delta T$ (red line) referenced with a linear curve (black line). **c** Crossectional profiles of $\delta T$ normalized by IR power 1 converted with the linear relation of $\delta n_\rho = 10^{-4}\delta T$. **d** Crossectional profiles of $\delta T$ normalized by IR power 1 converted with the nonlinear relation of $\delta n_\rho = 10^{-4}(0.98\delta T + 0.02\delta T^2)$.

**Supplementary Note 7: Single-exponential fitting for decay plots of temperature change.**

A single exponential fitting generally does not work well for evaluating thermal decay. We explain it by showing simulations of temperature change decay from two different micrometer-scale heat sources. Figure S7a represents a heat source with a Gaussian profile in the x-y plane and an exponential decay in the z-direction, which is similar to our MIP excitation, while Figure S7b shows a situation with a spherical heat source. The surrounding medium is water in both cases. The dots in the figures represent simulation results with 3D heat conduction calculation, showing thermal decay in microsecond time scale, while the solid lines are their fitting curves by single exponential functions. The calculation method of heat conduction is the same as that explained in Methods. It is clearly displayed that the fitting curves exhibit poor agreement with the decay plots, which strongly depend on the shape of the initial temperature distribution. This simulation suggests that single-exponential fitting is generally inappropriate for accurately determining thermal diffusivity. Therefore, we calculated 3D heat conduction to evaluate thermal diffusivity from our experimental data.

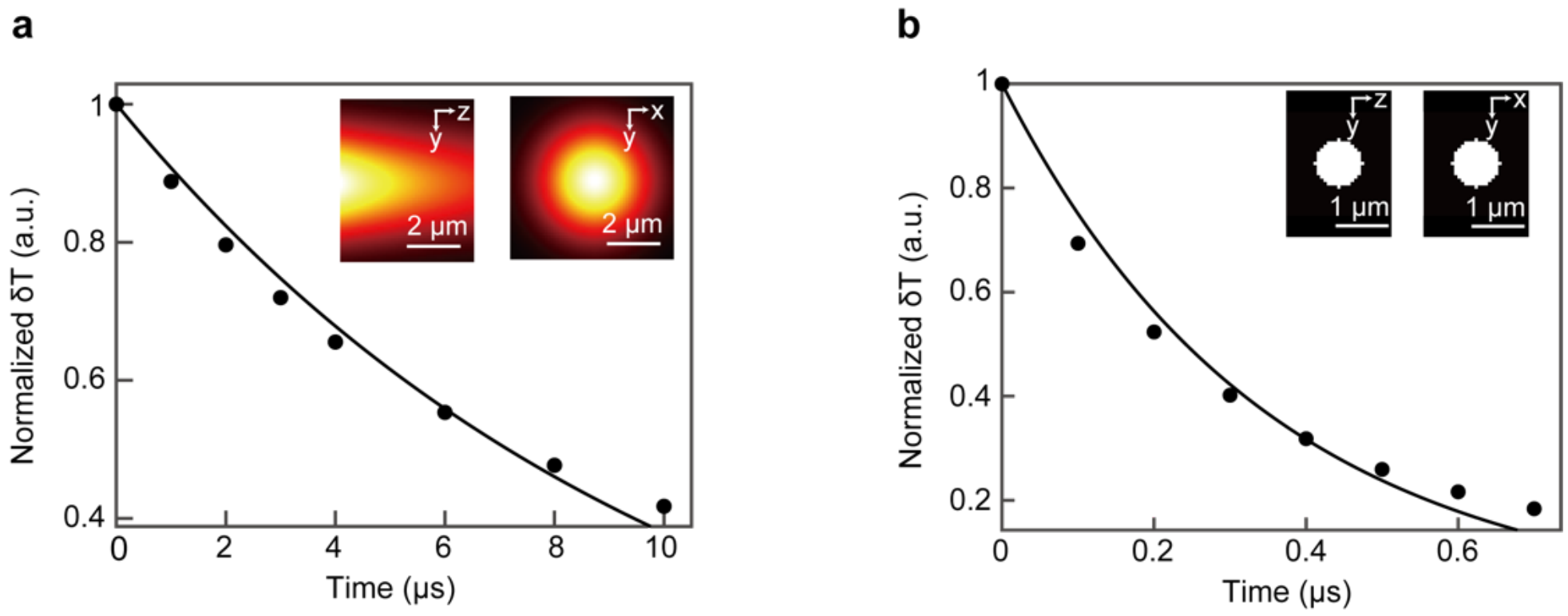

**Fig. S7 Single-exponential fitting of temperature change decay. a** A simulated temperature decay plot from an initial temperature distribution comprising a Gaussian profile with an FWHM of 2.7 μm in the x-y plane and an exponential function along the z-axis that attenuates in 6 μm. **b** A simulated temperature decay plot from an initial temperature distribution comprising a sphere with a diameter of 1 μm. Solid lines represent single-exponential fitting curves.

**Supplementary Note 8: Procedure of temperature measurement with fluorescent ratiometric sensing.**

Fluorescence ratiometric sensing[4] is a technique that enables measuring temperature by taking a ratio of separately measured fluorescence intensities at different wavelength bands, where temperature dependences are different. In this experiment, we take a signal with less dependence on temperature at 500-520 nm, and that with strong temperature dependence at 550-600 nm. The captured fluorescence images of a cell are presented in Fig. S8a. Figure S8b shows time variations of fluorescence intensities with continuous heating at a region indicated in the dotted red box shown in Fig. S8a. To extract temperature from the measured data, we made a calibration curve by measuring FPT signals in cells (5 cells) at different temperatures. We used a heating plate to control the temperature of a culture medium (TP-CHSQ-C, Tokai-Hit). Here, we assume the cell's temperature is equivalent to that of the surrounding culture medium. Figure S8c shows the calibration curve we obtained from the averaged data. Figure S8d shows the time variation of temperature derived from the measured data (Fig. S8b) using the calibration curve. Since the fluorescence measurements at different wavelengths were not simultaneously captured, the fluorescence intensities could suffer from the FPT's photobleaching, causing a slight variation in the fluorescence ratio. To remove this effect, we set the temperature at the time before heating becomes equivalent to that of the heater.

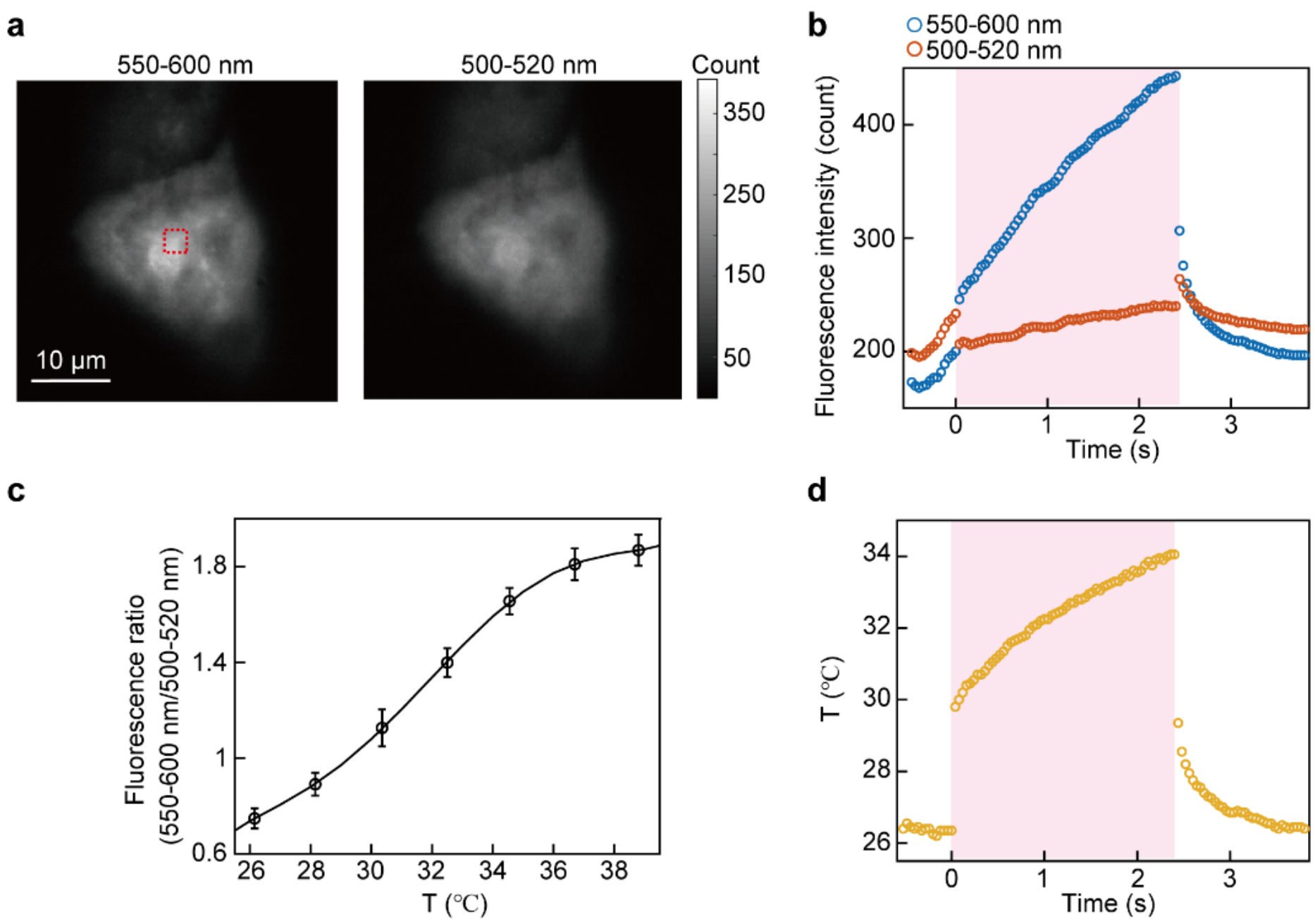

**Fig. S8 Temperature measurement with fluorescent ratiometric sensing. a** Fluorescence images of a cell measured at the wavelength bands of 550-600 nm (left) and 500-520 nm (right). **b** Time variations of fluorescence intensities averaged in the dotted red area in **a** (blue: 550-600 nm, orange: 500-520 nm). **c** A calibration curve of temperature to fluorescence ratio between the two wavelength bands. **d** Time variation of temperature derived from **b** using the calibration curve shown in **c**.

**Supplementary Note 9: Timing chart of MIP-ODT and fluorescence imaging for heat-induced intracellular temperature change measurements.**

Figures S9a and S9b show the timing chart of MIP-ODT thermometry and fluorescent nanothermometry, respectively.

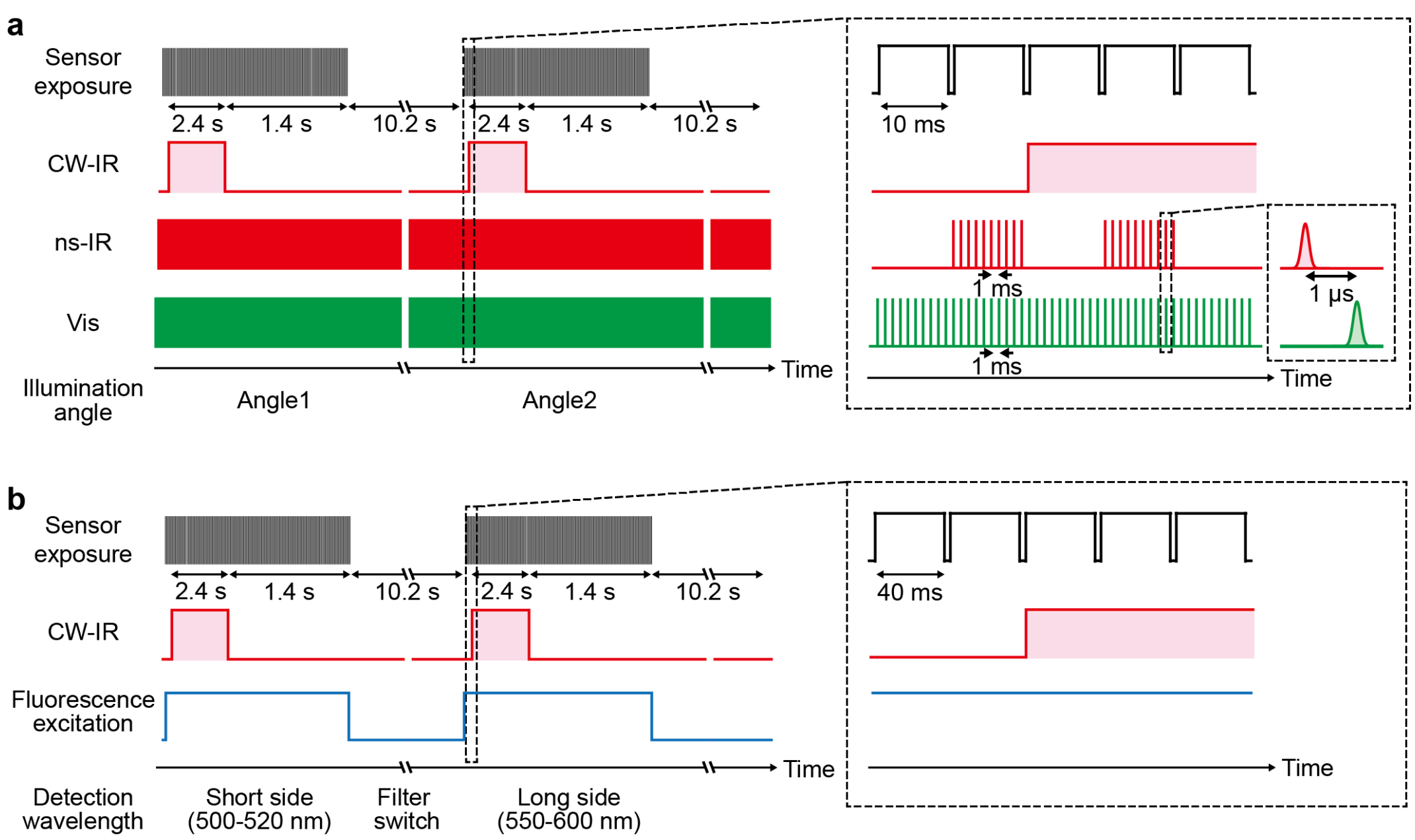

**Fig. S9 Timing chart for heat-induced temperature change measurement. a** Label-free MIP-ODT thermometry with dual-heating method. **b** Fluorescent nanothermometry with FPT.

**Supplementary Note 10: Validation of temperature measurement capability of the dual-heating MIP-ODT thermometer.**

To verify the temperature measurement capability of the dual-heating method with MIP-ODT, we conducted a well-controlled experiment using a well-characterized sample. We measured the temporal evolution of temperature variations in a water sample under seconds-long continuous heating with linearly increasing the heating power, as shown in Fig. S10. Given the rapid thermal diffusion in water, the temperature is expected to reach a steady state within 100 μs, ensuring that the temperature response closely follows the applied heating power on millisecond-to-second timescales. The black data points represent the temperature response measured using the dual-heating method, with temperature determined following the procedure described in the "Calibration curve of $\delta n_\rho$ on ΔT" in Methods. The orange line shows the temporal evolution of temperature obtained via conventional single-heating MIP-ODT, utilizing the established linear relationship between Δn and ΔT ($\Delta n = \frac{dn}{dT}\Delta T{\sim}10^{-4} \times \Delta T$). It is important to note that no thermophoresis effect is present in this measurement, as the sample consists of pure water. The measurement accurately captured the expected temperature response of the water sample, demonstrating a linear increase in temperature over time. There is excellent agreement between the conventional single heating and the dual-heating methods, confirming that temperature changes can reliably be measured through the temperature dependence of dn/dT. Consequently, the dual-heating method effectively serves as a robust reference thermometer in comparison with fluorescent thermometers.

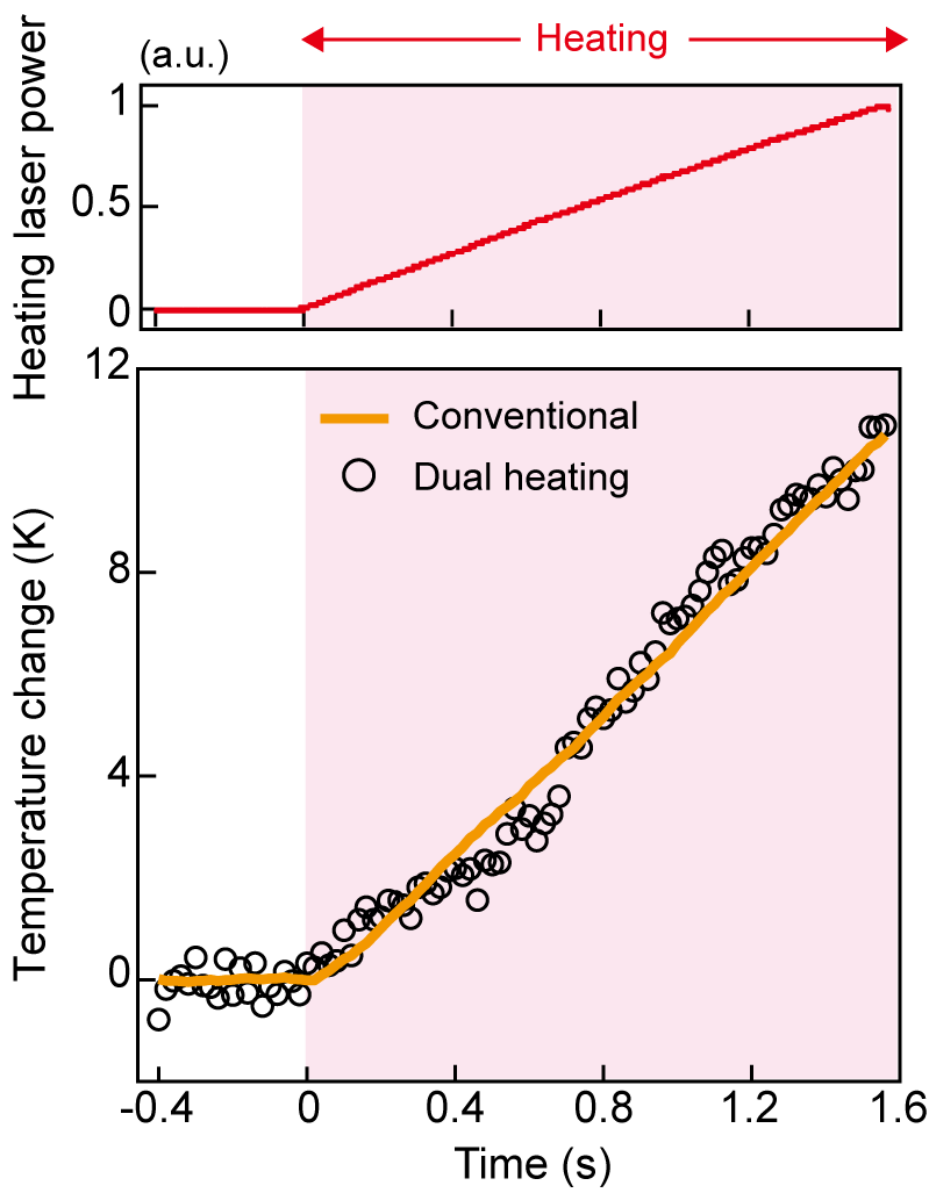

**Fig. S10 Validation of temperature measurement capability of the MIP-ODT thermometer using the dual-heating method.** The red line represents the dynamically changing heating laser power. The orange line and black

circles show temperatures obtained with the conventional single-heating and dual-heating methods using MIP-ODT, respectively. In the dual-heating method, temperature values were determined following the procedure described in the "Calibration curve of $\delta n_\rho$ on $\Delta T$" section in Methods.

**Supplementary Note 11: Change in refractive index during continuous heating.**

We measured the time variation of $\Delta n$ inside and outside a COS7 cell (Fig. S11a) using a heating duration of 0.25 s by subtracting the RI image captured right before heating (t = 0) from those obtained afterward. Figures S11b and S11c illustrate the $\Delta n$ map inside and outside the cell at t = 0.01, 0.25, 0.26, and 0.28 s, respectively. By comparing the results obtained at 0.01 s (the beginning of heating) and 0.25 s (the end of heating), we see a decreased $\Delta n$ at 0.25 s, specifically around the center of the heated area inside the cell, which persists even after the heating ends (see t = 0.26 and 0.28 s). Note that the color scales are displayed with negative values. Figure S11d represents $\Delta n$ time variations observed inside and outside the cell. The $\Delta n$ outside the cell (black squares) shows a rectangular time variation with instantaneous rise and fall, which is consistent with the water-like heat conduction. On the other hand, a gradual change in $\Delta n$ is observed inside the cell. We also observed site-specific different time variations at the nucleus and cytoplasm (see Supplementary Note 12). These results indicate that slow changes in $\Delta n_\sigma$ appear inside the cell during the seconds-long heating as expected. However, only the $\Delta n$ measurement cannot exclude the possibility of a slowly varying $\Delta n_\rho$. We introduce dual-heating method to solve this issue in the main text. There is another important insight obtained from Fig. S11d. At the beginning of the heating, $\Delta n$ instantly changes by a certain amount and then slowly evolves afterward. This implies that $\Delta n_\rho$ is the dominant factor in the initial short time of the heating. It is confirmed by evaluating a portion of the slow variation in $\Delta n$ being only 4.9% at 5 ms (half the frame interval) after the beginning of heating, as illustrated in Fig. S11d.

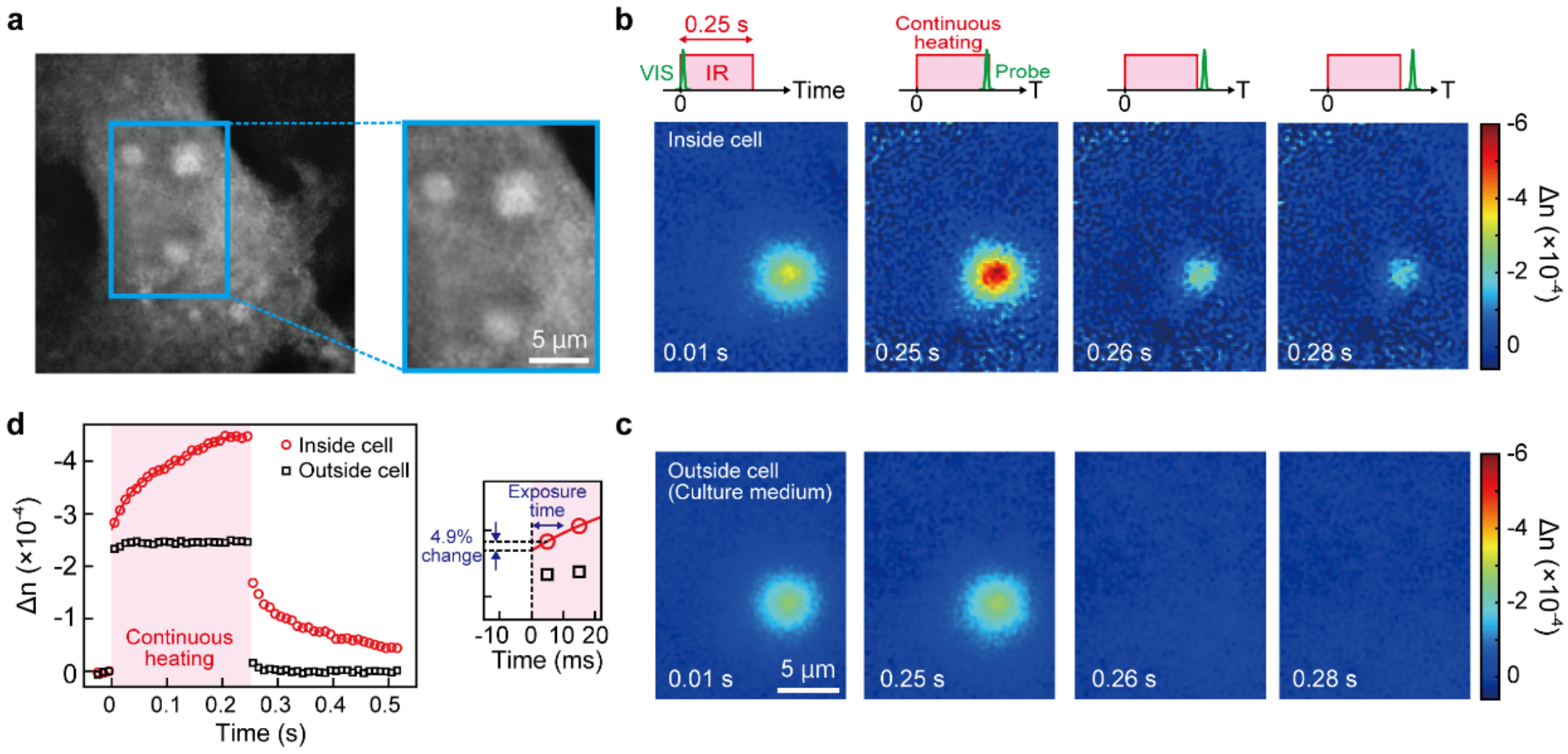

**Fig. S11 Refractive index change (Δn) measurement with continuous heating. a** An RI image of a COS7 cell (left) and a zoom-in around the heating area (right). **b** Δn images inside the cell at 0.01, 0.25, 0.26, and 0.28 s after the beginning of the continuous heating. **c** Δn images outside the cell at the same time frames as **b**. **d** Time variation of Δn inside (red) and outside (black) the cell. There is only a culture medium outside the cell. The small right panel

shows a zoom-in around the beginning of the heating. The portion of the slow variation at 5 ms is evaluated as 4.9% to the Δn value at 0 ms, confirming the $\Delta n_\rho$ is dominant at the beginning of the heating (~5ms).

**Supplementary Note 12: Site-specific intracellular time variation of $\Delta n$ induced by continuous heating.**

Figure S12 shows the site-specific time variations of $\Delta n$ within a COS7 cell, particularly at the nucleus and cytoplasm. Interestingly, the time variations are significantly different. The $\Delta n$ increases during heating in the nucleus, while it decreases in the cytoplasm. Further details regarding the site-specific $\Delta n$ time variations are discussed in another paper[5].

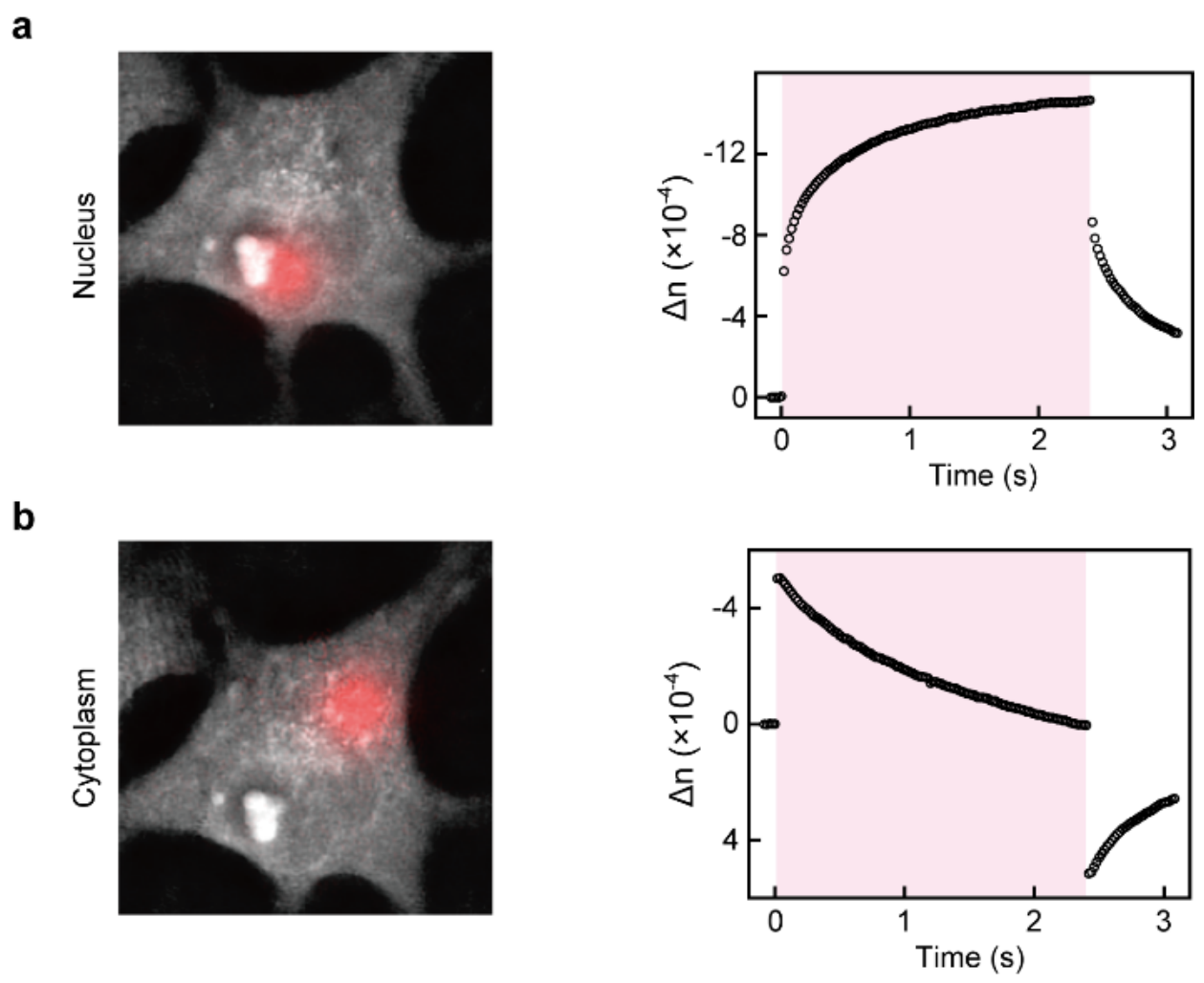

**Fig. S12 Site-specific time variations of $\Delta n$ in a cell induced by continuous heating.** RI images of a COS7 cell with the heating spots, **a** in the nucleus and **b** in the cytoplasm, displayed in red (left), and time variations of $\Delta n$ around the center of the heating spot (right).

**Supplementary Note 13: Frame interpolation method.**

We used the frame interpolation method in dual-heating conditions to cancel the effect of the slowly varying component ($\Delta n_\sigma$) in $\Delta n$ for precise evaluation of $\delta n_\rho$ by subtracting the values obtained with impulsive heating ON and OFF states. Figure S13a illustrates the method. To verify this method, we apply it to the data shown in Fig. S11d, which represents a slowly varying $\Delta n$ induced by continuous heating, which is shown as red circles in Fig. S13b. We plot interpolated values as black crosses, which are the averages of data points before ($\Delta n\ (t - 0.01\ s)$) and after ($\Delta n\ (t + 0.01\ s)$). Note that 0.01 s is the frame interval of our measurement. Blue squares in Figure S13b show the difference between $\Delta n(t)$ (red circles) and the interpolated values (black crosses). The STD of the difference is $2.9\times10^{-6}$, which is less than 1% of the $\Delta n$ values, showing this method largely reduces the slowly varying effect in $\Delta n$. While a linear frame interpolation was employed here, we used nonlinear frame interpolations with polynomial fitting for analyzing data shown in Figs. 3 and S16 to adapt various temporal evolution of $\Delta n$.

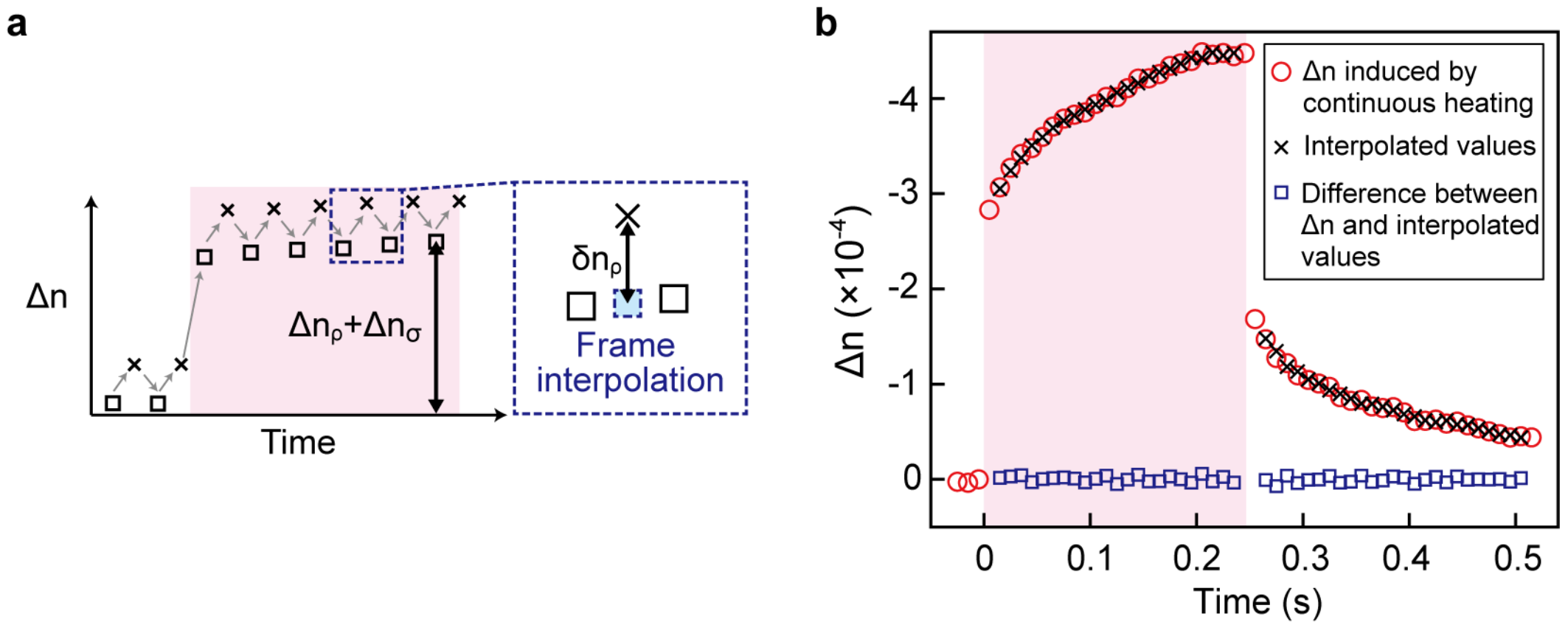

**Fig. S13 Frame interpolation method. a** Illustrative depiction of the method. **b** Validation of canceling a slowly varying effect by the frame interpolation method. The red circles present $\Delta n$ inside a cell during continuous heating. The black crosses show the interpolated values $(\Delta n\ (t+0.01\ s)+\Delta n\ (t-0.01\ s))/2$, and the blue squares exhibit the difference between $\Delta n$ and the interpolated values.

**Supplementary Note 14: Minor correction of $\Delta T$ by eliminating a slowly varying effect on dn/dT caused by a change in dry mass concentration.**

We derived $\Delta T$ by temperature dependence of dn/dT, but a change in dry mass concentration ($\sigma$) also minorly affect. To see this effect, we first break the slowly varying component of $\Delta n$ into $\Delta n_\rho+\Delta n_\sigma$ and see $\Delta n_\sigma$ is the dominant factor. As shown with the black squares in Fig. 3b, $\Delta n$ varies over ~$8\times10^{-4}$ (from $-4\times10^{-4}$ to $4\times10^{-4}$) during continuous heating. Black circles in Fig. S14 show the subtraction between impulsive heating on and off, leading to $\Delta T$ evaluated through $\delta n_\rho$. The time variation of $\Delta T$ shows almost constant rise of 5K with a slight increase by ~0.5K indicated by red arrows. If we assume this slight temperature increase was due to $\Delta n_\rho$, the expected value would be ~$5\times10^{-5}$ calculating with the thermo-optic coefficient of ~$1\times10^{-4}$. Since this value is more than an order of magnitude smaller than $\Delta n$ variation of ~$8\times10^{-4}$, we can conclude that $\Delta n_\sigma$ is the dominant factor of the slowly varying component of $\Delta n$.

Next, we evaluate dn/dT variation caused by $\Delta n_\sigma$. The dn/dT of homogeneous solids and liquids can be written as dn/dT$\propto (n^2-1)\times(n^2+2)/6n^2$ as reported in literature[6]. The Taylor expansion around n leads to

$$\left(\frac{dn}{dT}\right)_{n+\Delta}=\left(\frac{dn}{dT}\right)_{n}+\beta\Delta \qquad \text{(Eq. S1)},$$

Where $\Delta$ is a slight deviation of n, and $\beta$ is a constant value. The cellular RI can be written by the summation of RI of water ($n_{\mathrm{water}}$) and RI increment due to dry mass concentration ($n_\sigma$)[7], $n_{\mathrm{cell}}=n_{\mathrm{water}}+n_\sigma$. Therefore, the intracellular dn/dT may be written as

$$\left(\frac{dn}{dT}\right)_{\mathrm{cell}}=\left(\frac{dn}{dT}\right)_{\mathrm{water}}+\beta n_\sigma \qquad \text{(Eq. S2)}$$

Since the intracellular dn/dT is 1.1-fold larger than that of water, as shown in Supplementary Note 15, Eq. S2 can be expressed as

$$\left(\frac{dn}{dT}\right)_{\text{cell}} = 1.1 \times \left(\frac{dn}{dT}\right)_{\text{water}} \quad \text{(Eq. S3).}$$

From Eqs. S2 and S3, we obtain

$$\beta = 0.1 \times (dn/dT)_{\text{water}}/n_{\sigma} \quad \text{(Eq. S4).}$$

Now, we consider the time variation of dn/dT during continuous heating. Since time variation of RI can be written as $n_{\text{cell(heating)}} = n_{\text{cell}} + \Delta n_{\sigma}(t)$, dn/dT can be written as

$$\left(\frac{dn}{dT}\right)_{\text{cell(heating)}} = \left(\frac{dn}{dT}\right)_{\text{cell}} + \beta \Delta n_{\sigma}(t)$$

By substituting Eqs. S3 and S4, we obtain

$$\left(\frac{dn}{dT}\right)_{\text{cell(heating)}} = \left(\frac{dn}{dT}\right)_{\text{cell}} \left(1 + \frac{0.1}{1.1} \times \frac{\Delta n_{\sigma}(t)}{n_{\sigma}}\right) \quad \text{(Eq. S5)}$$

The Equation S5 suggests that the ratio of the intracellular dn/dT with and without heating can be derived from the ratio of the dry mass concentration with and without heating ($\Delta n_{\sigma}(t)/n_{\sigma}$). Note that we can experimentally evaluate $n_{c_p}$ from the RI image without heating.

The blue solid line in Fig. S14 represents the rate of time variation of dn/dT induced by $\Delta n_{\sigma}(t)$. The red circles show a time variation of $\Delta T$ after eliminating this effect, which is obtained by dividing $\Delta T$ before correction (black circles) by the time variation of dn/dT (blue solid line).

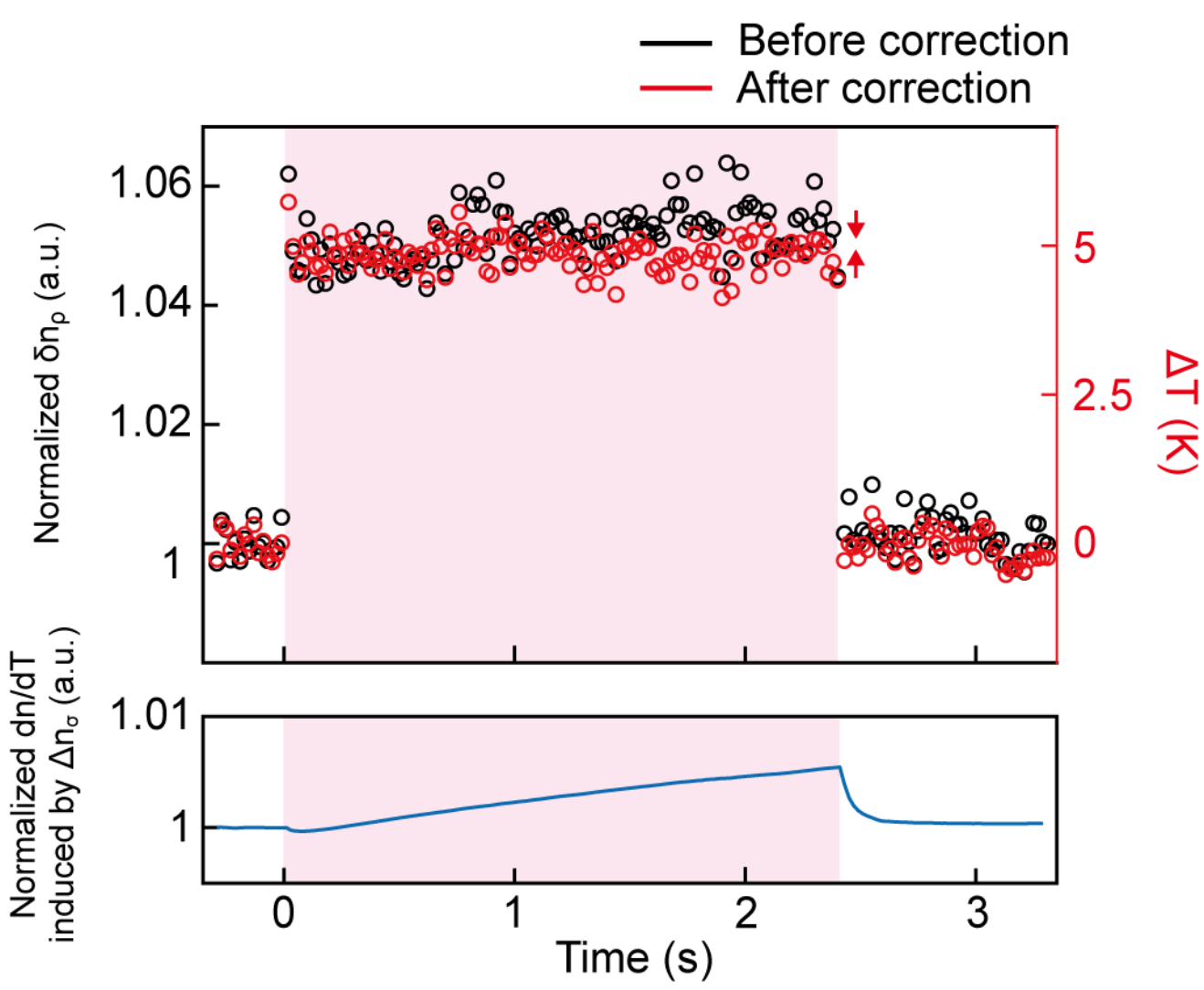

**Fig. S14 Correction of $\Delta T$ to eliminate minor effects arising from $\Delta\sigma$.** The top panel shows normalized $\delta n_p$ and corresponding time variation of $\Delta T$ during continuous heating. The black and red circles represent measured data before and after the correction. The bottom panel shows the time variation of dn/dT induced by $\Delta\sigma$ normalized by the value before heating (blue line).

**Supplementary Note 15: Experimental validation of intracellular thermo-optic coefficient.**

We experimentally validate the intracellular thermo-optic coefficient, dn/dT, by comparing the z-axis integration of heat-induced RI change of a cell to that of water. The RI change right after heating, which is before being affected by thermal diffusion, can be expressed as

$$\delta n_\rho = \frac{dn}{dT}\frac{\phi}{c\rho},$$

where $\phi$, $c$, and $\rho$ represent the total absorbed energy of IR photons, the specific heat capacity, and the density of the sample. Since the value of the specific heat capacity times density $c\rho$ of a cell (4,210 J/K $cm^3$) is very close to that of water (4,180 J/K $cm^3$), the difference in the evaluated value between water and the cell can be attributed to the difference in dn/dT under the condition that the same amount of IR light is absorbed. To ensure this condition and avoid the effect of thermal diffusion, we employed the ns pump-probe imaging method with IR pulses at a wavenumber of 3,150 $cm^{-1}$ and VIS pulses with a delay set to 200 ns. Our experimental comparison revealed that the intracellular dn/dT was 1.1-fold larger than that of water. This difference could arise from the RI difference between a cell and water due to the dry mass concentration because dn/dT is known to depend on the RI with a relation of dn/dT$\propto (n^2-1)\times(n^2+2)/6n$ as reported in literature[6]. We also confirmed that the experimentally obtained 1.1-fold larger dn/dT of a cell than water showed a good agreement with a theoretical estimation with this equation using a cellular RI of 1.36, which was derived from RI increment due to proteins (0.19 l/mg)[1] and dry mass concentration in a typical cell (~0.15 g/ml)[2].

**Supplementary Note 16: Calibration curve of $\delta n_\rho$ on $\Delta T$.**

Figure S15 shows an experimentally measured linear dependence of normalized $\delta n_\rho$ on $\Delta T$.

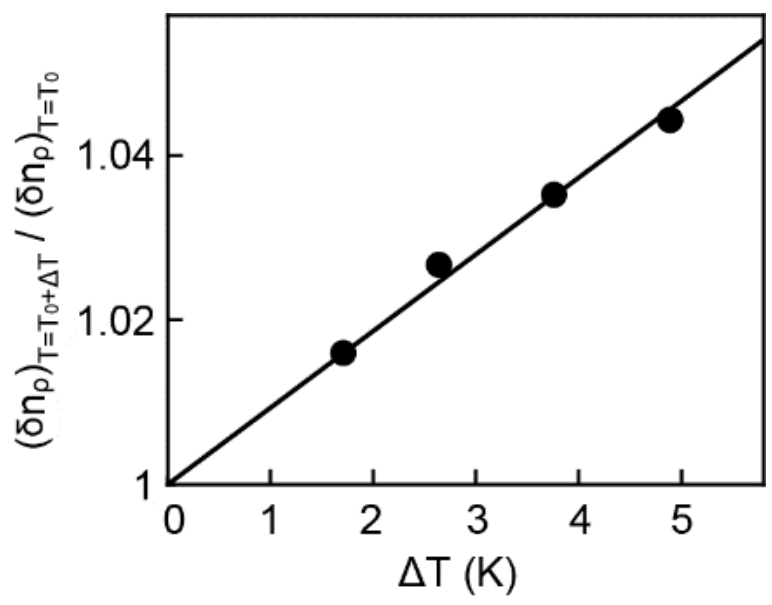

**Fig. S15** An experimentally measured linear dependence of normalized $\delta n_\rho$ on $\Delta T$.

**Supplementary Note 17: Measured $\Delta T$ for a cell showing a different time variation in $\Delta n_\sigma$ from that represented in Fig. 3.**

Figure S16 illustrates the time variation of $\Delta T$ for a cell that exhibits a distinct characteristic of $\Delta n_\sigma$ variation as compared to that displayed in Fig. 3. The time variation of $\Delta T$ shows good agreement with that shown in Fig. 3, verifying that measured $\Delta T$ does not majorly depend on $\Delta n_\sigma$. Note that $\Delta n_\sigma$ slightly affects $\Delta T$ as explained in Supplementary Note 14.

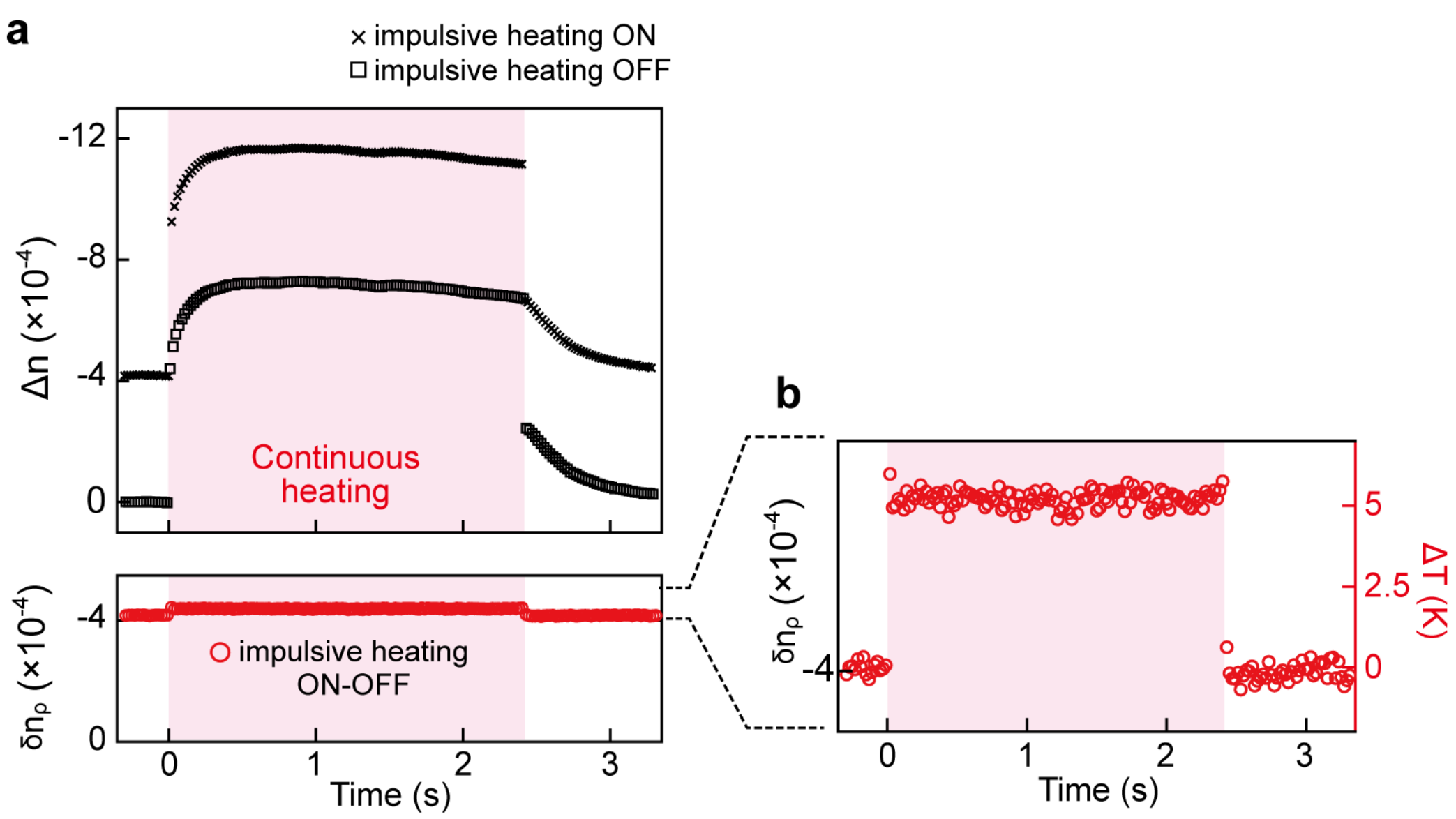

**Fig. S16 $\Delta T$ measurement for a cell that exhibits a different temporal evolution in $\Delta n_{c_p}$ from that represented in Fig. 3. a** A temporal evolution of $\Delta n$ under continuous heating together with (black crosses) and without (black squares) the impulsive heating, and their difference, $\delta n_\rho$ (red circles) **b** $\Delta T$ evolution induced by continuous heating measured by MIP-ODT.

**Supplementary Note 18: Evidence from an independent study excluding artifacts in fluorescent nanothermometry.**

To rigorously exclude the possibility that the slowly varying signal observed in Fig. 3 originates from artifacts intrinsic to fluorescent nanothermometry, we refer to independent validation experiments performed by our co-authors[8]. First, direct IR excitation of intracellular biomolecules was ruled out. In ref. S8, the slowly varying signal persisted even when IR illumination was restricted to the extracellular medium. Under these conditions, intracellular temperature increases occurred solely through inward heat conduction, without direct optical absorption inside the cell. This finding eliminates intracellular IR absorption as the origin of the slow signal.

Second, the intrinsic temperature-response kinetics of the fluorescent probe were evaluated. The response time of FPT in the cellular environment was measured to be 13 ms, which is several orders of magnitude shorter than the seconds-scale evolution of the slowly varying component. The probe's intrinsic response dynamics therefore cannot account for the observed slowly varying behavior.

Third, probe diffusion was quantitatively evaluated. The diffusion coefficient of FPT was measured to be 0.13 μm$^2$ s$^{-1}$. Within the intrinsic response time of 13 ms, the expected diffusive displacement, calculated as $\sqrt{2Dt}$, is ~60 nm. This displacement is well below the optical diffraction limit and substantially smaller than the micrometer-scale dimensions of the heated region. Under these conditions, diffusion can only spatially redistribute the probes and cannot distort the observed temperature distribution or generate delayed temporal dynamics in the fluorescence readout. Therefore, it is highly unlikely that the measured slow signal originates from probe diffusion.

**Supplementary Note 19: Effect of metabolic inhibition on the slowly varying fluorescence signal.**

To determine whether the slowly varying component observed in Fig. 3 is associated with metabolic activities, we examined its temporal behavior under pharmacological inhibition of the major metabolic energy pathways. Figure S17 compares temperature responses measured by FPT in untreated control cells and in cells treated with inhibitors of glycolysis and oxidative phosphorylation. Glycolysis was suppressed using 2-deoxy-D-glucose (2-DG, 10 mM), whereas oxidative phosphorylation was inhibited by rotenone and antimycin A (1 μM each). Despite effective metabolic suppression, the slowly varying component persisted and exhibited temporal dynamics comparable to those observed in control cells. This result indicates that the slowly varying signal does not primarily arise from the major metabolic pathways.

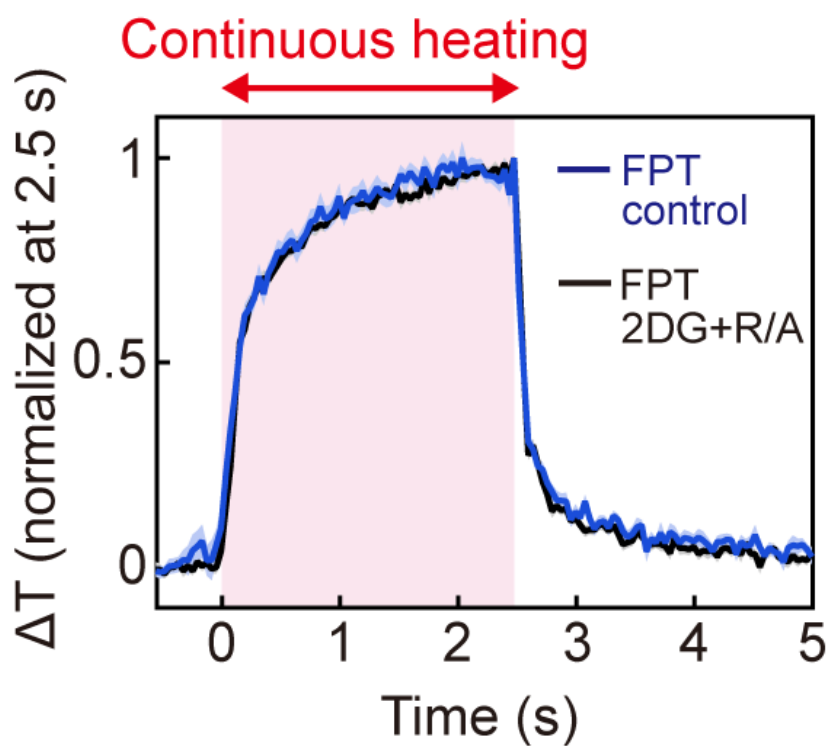

**Fig. S17 FPT response under metabolic inhibition.** Temporal evolution of the temperature signal measured by FPT during continuous IR heating in control cells and in cells treated with inhibitors of glycolysis and oxidative phosphorylation. Shaded regions indicate the standard error of the mean (s.e.m.), with n = 30 independent measurements.